\documentclass[prb,aps,twocolumn,superscriptaddress,showpacs]{revtex4-1}
\usepackage{amsmath}
\usepackage{amssymb}
\usepackage{bbold}
\usepackage{mathrsfs}
\usepackage{bm}
\usepackage{graphicx}
\usepackage[utf8]{inputenc}
\usepackage[english]{babel}
\usepackage{dsfont}

\begin{document}
\title{A functional calculus for the magnetization dynamics}
\author{Julien Tranchida}
\email{julien.tranchida@cea.fr}
\affiliation{CEA DAM/Le Ripault, BP 16, F-37260, Monts, FRANCE}
\affiliation{CNRS-Laboratoire de Mathématiques et Physique Théorique (UMR 7350), Fédération de Recherche "Denis Poisson" (FR2964), Département de Physique, Université de Tours, Parc de Grandmont, F-37200, Tours, FRANCE}
\author{Pascal Thibaudeau}
\email{pascal.thibaudeau@cea.fr}
\affiliation{CEA DAM/Le Ripault, BP 16, F-37260, Monts, FRANCE}
\author{Stam Nicolis}
\email{stam.nicolis@lmpt.univ-tours.fr}
\affiliation{CNRS-Laboratoire de Mathématiques et Physique Théorique (UMR 7350), Fédération de Recherche "Denis Poisson" (FR2964), Département de Physique, Université de Tours, Parc de Grandmont, F-37200, Tours, FRANCE}

\date{\today}                                           

\begin{abstract}
A functional calculus approach is applied to the derivation of evolution equations for the moments of the magnetization dynamics of systems subject to stochastic fields. It allows us to derive a general framework for obtaining the master equation for the stochastic magnetization dynamics, that is applied to both, Markovian and non-Markovian dynamics. The formalism is applied for studying different kinds of interactions, that are of practical relevance and hierarchies of evolution equations for the moments of the distribution of  the magnetization are obtained. 
In each case, assumptions are spelled out, in order to close the hierarchies. These closure assumptions are tested by extensive numerical studies, that probe the validity of Gaussian or non--Gaussian closure Ansätze. 
\end{abstract}


\maketitle
\section{Introduction}\label{Intro}
Thermal fluctuations of the magnetization are a significant factor for the operating conditions of magnetic devices\cite{evans2012thermally,suhl2007relaxation}. To describe them well is quite challenging, even in cases where the thermal effects are, not an inconvenience, but essential for eliciting the desired magnetic response~\cite{ostler2012ultrafast,thiele2003ferh}, and the development of appropriate computational methods has a long history~\cite{neel1953thermoremanent,brown1979thermal,coffey2012thermal}. 

A textbook approach for the description of thermal fluctuations is the stochastic calculus~\cite{gardiner1985stochastic,van1992stochastic}: the fluctuations are described by a thermal bath, interacting with the magnetic degrees of freedom, namely, spins, and the quantities of interest are the correlation functions of the magnetization, deduced from numerical simulations~\cite{simon2014spin,tranchida_closing_2016}.

These correlation functions, in principle, define a measure on the space of spin configurations. This measure can be, either deduced from a Fokker--Planck equation~\cite{risken1984fokker,Mayergoyz:2009kl}, or, a Langevin equation. In the former case, this is a partial differential equation for the probability density, $P(\bm{\sigma},t)$, to find the magnetization vector $\bm{s}(t)$ in a state $\bm{\sigma}$ at time $t$; in the latter case, it is a partial differential equation for the magnetization, considered as a time--dependent field.

While at the level of a single spin, this approach can only describe transverse, but not longitudinal, damping effects \cite{evans2014atomistic,beaujouan2012anisotropic}, it has been shown~\cite{garanin1990dynamics,garanin1997fokker,kazantseva2008towards} that an appropriate averaging procedure over the bath can, in fact, describe longitudinal damping effects, that are typical in finite size magnetic grains. Thus, it may be a good starting point for developing models that incorporate the corrections to the mean field behavior of a single domain, beyond the effective medium approximation~\cite{bouchaud1989spontaneous,thibaudeau2015frequency}. Damping is responsible for the transfert of spin angular momentum from the magnetization to the environment and allows conversely energy to be pumped from the environment to the magnetization. Many different mechanisms for damping are already known that include spin-orbit coupling, lattice vibrations and spin-waves. At several levels, these mechanisms are limiting factors in the reduction of the remagnetization rate in magnetic recording devices. To better describe such damping effects, it's useful to refine the approach used to date for obtaining the evolution equations  towards equilibrium for the magnetization and its fluctuations. To this end a functional calculus approach~\cite{justin1989quantum,zinn2007phase,kleinert2009path} can be very efficient, and has been further developed recently~\cite{aron2014magnetization,moreno2015langevin}.

This approach has as starting point the functional integral over the bath degrees of freedom, $\bm{\eta}(t)$,
\begin{equation}
Z=\int\,[{\mathscr D}\bm{\eta}(t)]\,\rho(\bm{\eta}(t)) \label{bath_part_fun}
\end{equation}
The density, $\rho(\bm{\eta})$, is defined by its correlation functions, that are assumed to define a Gaussian process, that's completely given by  its two first moments: 
\begin{eqnarray}
\langle \eta_a(t) \rangle &=&0\nonumber \\
\langle \eta_a(t) \eta_b(t') \rangle &=&\delta_{ab}C(t-t') \nonumber
\label{General_noise}
\end{eqnarray}
with $C(t-t')$ a function, that, therefore depends not on both times, $t$ and $t'$, but only on their difference and describes the Markovian property and eventual deviations therefrom. All other correlation functions are expressed  using  Wick's theorem~\cite{justin1989quantum}. 

In the Markovian limit, the function $C(t-t')$ is ultra--local, namely,
\begin{equation}
C(t-t')=2D\delta(t-t')
\label{White_noise}
\end{equation}
where $D$ sets the scale of the bath fluctuations. This limit is relevant for cases when the auto--correlation time of the bath variables can be neglected. 

However, recent progress in magnetic devices has led to situations where this is no longer the case \cite{beaurepaire1996ultrafast,ounadjela2003spin}. Hence it is not only of theoretical, but also of practical interest, to develop tools for the quantitative description of baths with finite auto--correlation time\cite{fox1988fast,hanggi1995colored}. Examples are provided by the experimental study and simulations of extremely fast magnetic events. There it was found that a colored form for the noise, for which 
\begin{equation}
C(t-t')=\frac{D}{\tau}~{e}^{~ -\frac{|t-t'|}{\tau} }
\label{Colored_noise}
\end{equation}
can lead to good agreement between experiment and simulations \cite{bose2010correlation,miyazaki1998brownian,atxitia2009ultrafast}. In this expression, the auto--correlation time, $\tau$, describes the finite memory of the bath and, therefore, the inertial effects of its response, assuming isotropic in space.

That this expression is a reasonable generalization of the Markovian limit may be deduced from the fact that, in the limit $\tau\to 0$, the Markovian limit (also called ``white noise limit'') is recovered. The ``white noise limit'' may therefore be considered as the limiting case of colored noise for ``extremely short'' auto-correlation time \cite{tranchida2015colored}. What sets the scale of ``extremely short'' is at the heart of the subject and  depends on the detailed dynamics, that will be presented in the sections to follow.  

Having described the bath, we must now describe the degrees of freedom, whose dynamics is of interest, i.e. the spins. This dynamics is specified by a particular choice of the Langevin equation, the so--called stochastic form of the Landau-Lifschitz-Gilbert equation (sLLG). Up to a renormalization over the noise \cite{Mayergoyz:2009kl}, the sLLG equation of motion for each spin component $s_i$ can be written as follows~:
\begin{equation}
\frac{ds_i}{dt}=\frac{1}{1+\lambda^2}\,\epsilon_{ijk}\,s_k \left(\omega_j+\eta_j - \lambda\, \epsilon_{jlm}\,\omega_l\, s_m\right),
\label{sLLG}
\end{equation}
where the Einstein summation convention is adopted, and $\epsilon_{ijk}$ describes the Levi-Civita fully antisymetric pseudo-tensor. This equation describes purely transverse damping with a non-dimensional constant $\lambda$: properly integrated \cite{roma2014numerical,d2005geometrical}, it ensures that the norm of the spin remains constant, which one can normalize to unity, $\left|\bm{s} \right|=1$.

The vector $\bm{\omega}$ sets the precession frequency and in a Hamiltonian formalism is given  by the expression
\begin{equation}
\omega_i=-\frac{1}{\hbar}\frac{\partial{\mathscr H} }{\partial s_i},
\end{equation}
where ${\mathscr H}$ is the Hamiltonian of the system. Therefore, at equilibrium, for given ${\mathscr H}$, the ensemble average of the spin, $s$, along direction, $i$, $\langle s_i\rangle$, is given by the canonical average 
\begin{equation}
\label{ensemble_spin}
\left\langle s_i\right\rangle\equiv Z^{-1}\int\,[{\mathscr D}\bm{s}]\, s_i\,e^{-\beta{\mathscr H}({\bm s})}\equiv\int\,[{\mathscr D}\bm{s}]\,s_i\, P_\mathrm{eq}(\bm{s})
\end{equation} 
where $\beta=1/k_BT$ and the equilibrium distribution, $P_\mathrm{eq}(\bm{s})$, is given by the Gibbs expression
\begin{equation}
\label{Peq}
P_\mathrm{eq}(\bm{s})=\frac{e^{-\beta{\mathscr H}(\bm{s})}}{\int\,[{\mathscr D}\bm{s}]\,e^{-\beta{\mathscr H}(\bm{s})}}.
\end{equation}

In their seminal studies, Garanin \emph{et al.} \cite{garanin1990dynamics,garanin1997fokker} used this form of the equilibrium distribution to derive a Landau-Lifschitz-Bloch model from the  Fokker--Planck formalism, close to equilibrium. 

In this paper we do not assume the form of the equilibrium distribution, but we try to deduce its properties from the evolution of the off--equilibrium dynamics of equal--time correlation functions. To this end, we explore the consequences of closure schemes for the evolution equations. 

The plan is the following:

In section~\ref{Section_moments} we obtain the evolution equations for the equal--time 1-- and 2--point correlation functions for the spin components, taking into account different interaction Hamiltonians, namely Zeeman, anisotropy and exchange. We work in the mean field approximation and we use the results of appendices~\ref{App1} and~\ref{App2}. 

These equations are part of an open hierarchy. To solve them, we must impose closure conditions. 

In section~\ref{closure} we explore Gaussian, as well as non--Gaussian closure conditions, based on the theory of chaotic dynamical systems.  To test their validity we compare the results against those of a  ``reference model'',  studied within the framework of stochastic atomistic spin dynamics simulations. 

Our conclusions are presented in section~\ref{conclusions}. 

Technical details are the subject of the appendices. In particular,  in appendix~\ref{App3} we obtain, by functional methods,  a local form for the master equation, for the case of Ornstein--Uhlenbeck noise, in an expansion in the auto--correlation time of the noise, that's consistent with the symmetries of the problem. 

\section{Evolution equations for the correlators of the magnetization dynamics}\label{Section_moments}
In order to derive equations for the moments that capture the properties of   the magnetization dynamics, the probability $P(\bm{\sigma},t)$ to find the magnetization $\bm{s}(t)$ in a state $\bm{\sigma}$ at a time $t$ has to be properly defined. Within the functional calculus approach\cite{justin1989quantum}, this is realized by a path integral:
\begin{equation}
P({\bm\sigma},t)\equiv\int \left[ {\mathscr D}{\bm\eta}(t) \right] \rho\left({\bm\eta}(t)\right) {\bm\delta}\left( {\bm s}[{\bm\eta}(t)]-{\bm\sigma} \right)  \label{Proba_spin},
\end{equation}
where ${\bm s}[{\bm\eta}(t)]$ is a functional of the noise and ${\bm\delta}$, the functional $\delta-$distribution.
At equal times, any correlation function of the spin components is  given by
\begin{eqnarray}
 \langle F(\bm{s}(t)) \rangle &=&  \int d\bm{\sigma}F(\bm{\sigma})P(\bm{\sigma},t). \label{AverageM}
\end{eqnarray}
Its   time derivative can be constructed from elementary building blocks, that are the multi--component correlation functions as follows:
\begin{eqnarray}
\frac{d}{dt}  \langle s_{i_1} (t) \cdots s_{i_k}(t) \rangle &=& \int d\bm{\sigma}\, \sigma_{i_1}\cdots \sigma_{i_k}\, \frac{\partial P(\bm{\sigma},t)}{\partial t}.\label{spin-time-derivatives}
\end{eqnarray}
These expressions become even more explicit upon replacing $P(\bm{\sigma},t)$ by the expression in eq.(\ref{Proba_spin}) and by performing the functional integral  over the noise. For $P(\bm{\sigma},t)$, this produces an integro--differential master equation--that will become a Fokker--Planck equation in an appropriate limit. Details are given in appendix~\ref{App1}. Formally this can always be written as a continuity equation
\begin{equation}
\frac{\partial P(\bm{\sigma},t)}{\partial t}=-\frac{\partial J_i({\bm\sigma},t)}{\partial\sigma_i},
\end{equation}
with the divergence of the probability flow ${\bm J}\left(\bm{\sigma},t \right)$, obtained from eq.(\ref{Probability_flow}). Equation (\ref{spin-time-derivatives}) can then be simplified by partial integration, where the surface terms can be dropped, since the manifold, described by the spin variables, is a sphere--i.e. does not have a boundary. Whether defects on the manifold could contribute is very interesting, but beyond the scope of the present investigation. Thus any moment of the spin variables can be computed from  this expression as
\begin{eqnarray}
\frac{d}{dt} \langle s_{i_1} (t) \cdots s_{i_k}(t) \rangle &=& \int d\bm{\sigma}\, J_{j}\left(\bm{\sigma},t \right) \frac{\partial}{\partial \sigma_j} \left(\sigma_{i_1}\cdots \sigma_{i_k} \right).
\end{eqnarray} 
For Markovian dynamics, the probability flow ${\bm J}$ is given by eq.(\ref{Probability_flow_White}). If we rewrite eq.(\ref{sLLG}) as
\begin{equation}
\frac{ds_i}{dt}=A_i(\bm{s})+e_{ij}(\bm{s})\eta_j(t),
\end{equation}
 the evolution equations of the first and second moments become
\begin{widetext}
\begin{eqnarray}
\label{eveq12}
\frac{d\langle s_i\rangle}{dt}&=&\langle A_i(\bm{s})\rangle+D\Big\langle \frac{\partial(e_{il}(\bm{s})e_{ml}(\bm{s}))}{\partial s_m}\Big\rangle\\
\frac{d\langle s_is_j\rangle}{dt}&=&\langle A_i(\bm{s})s_j\rangle+\langle A_j(\bm{s})s_i\rangle+D\left(\Big\langle \frac{\partial(s_ie_{jl}(\bm{s})e_{ml}(\bm{s}))}{\partial s_m}\Big\rangle+\Big\langle \frac{\partial(s_je_{il}(\bm{s})e_{ml}(\bm{s}))}{\partial s_m}\Big\rangle\right)
\end{eqnarray} 
\end{widetext}
in the white noise limit. 
The RHS of these equations can be expressed as follows, where the exponent $M=0$ or 1:
\begin{widetext}
\begin{eqnarray}
\label{RHSeveq12}
\langle A_is_l^M\rangle&=&-\frac{1}{\hbar(1+\lambda^2)}\left(\epsilon_{ijk}\Big\langle\frac{\partial{\mathscr H}}{\partial s_j}s_ks_l^M\Big\rangle+\lambda\left(\Big\langle \frac{\partial{\mathscr H}}{\partial s_i}s_js_js_l^M\Big\rangle-\Big\langle\frac{\partial{\mathscr H}}{\partial s_j}s_js_is_l^M\Big\rangle\right)\right)\nonumber\\
\Big\langle \frac{\partial(e_{il}(\bm{s})e_{ml}(\bm{s}))}{\partial s_m}\Big\rangle&=&-\frac{2}{(1+\lambda^2)^2}\langle s_i\rangle\nonumber\\
\Big\langle \frac{\partial(s_je_{il}(\bm{s})e_{ml}(\bm{s}))}{\partial s_m}\Big\rangle&=&\frac{1}{(1+\lambda^2)^2}\left(\delta_{ij}\langle s_ks_k\rangle-3\langle s_is_j\rangle\right)\nonumber
\end{eqnarray}
\end{widetext}
We shall call the terms proportional to $D$ in eq.~(\ref{eveq12}) ``longitudinal'', because they affect the norm of the average magnetization--whereas the other terms we shall call ``transverse'', since $A_i(\bm{s})$ does, of course, affect the components transverse to the direction of the instantaneous magnetization; however it's important to keep in mind that its average, $\langle A_i(\bm{s})\rangle$, may not be purely transverse. 

In any event, these expressions highlight that the terms proportional to the amplitude of the noise, $D$, 
 are independent of the particular choice of a local Hamiltonian, because it is not a part of the vielbein $e_{ij}({\bm{s}})$, whereas the ``transverse'' terms explicitly depend on this choice. 

For a single atomic spin, it will be useful to start with a ultra--local expression for the Hamiltonian, ${\mathscr H}$, consisting of a Zeeman term and an anisotropy term:
\begin{equation}
{\mathscr H}=\underbrace{-g \mu_B s_i B_i}_{\rm Zeeman} -\underbrace{\frac{K_a}{2} \left( \left(n_is_i\right)^2-1 \right)}_{\rm Anisotropy}. \label{Hamiltonian}
\end{equation}
In the Zeeman energy, $g$ is the the gyromagnetic ratio, $\mu_B$ the Bohr magneton and ${\bm{B}}$ the external magnetic induction. The anisotropic energy term describes a uniform uniaxial anisotropy, defined by an easy-axis ${\bm{n}}$ and  intensity $K_a$. 

Let us consider, for the moment, only the Zeeman contribution. Even if the external magnetic field, $\bm{\omega}$, may, in general, depend on time, it is assumed to be independent of the noise and therefore can be taken out of any noise average.

The corresponding expressions for the first and second spin moments, therefore, are 
\begin{widetext}
\begin{eqnarray}
\frac{d\langle s_i\rangle}{dt} &=& \frac{1}{1+\lambda^2}\epsilon_{ijk} \Big(\omega_j \,\langle s_k\rangle  -\lambda \,\epsilon_{jlm}\,  \omega_l\, \langle s_k s_m \rangle \Big)  -\frac{2D}{\left(1+\lambda^2  \right)^2}\, \langle s_i\rangle,  \label{Average_Zeeman1} \\
\frac{d \langle s_i s_j \rangle}{dt} &=& \frac{1}{1+\lambda^2}\epsilon_{ikl} \Big( \omega_k\langle s_j s_l\rangle -\lambda\, \epsilon_{kmn}\, \omega_m\, \langle s_j s_l s_n\rangle \Big)  +\frac{D}{\left( 1+\lambda^2\right)^2} \Big( \delta_{ij}\langle s_n s_n \rangle - 3\langle s_i s_j \rangle \Big) \nonumber\\
&&+\left(i \leftrightarrow j \right). \label{Average_Zeeman2}
\end{eqnarray}
\end{widetext}
It is striking that these equations are  very similar to those obtained by Garanin \emph{et al.} \cite{garanin1990dynamics}. 

We observe that they are  not closed. Indeed, the RHS of eq.(\ref{Average_Zeeman2}) contains three--point moments $\langle s_j s_l s_n\rangle$, which are not defined yet. 

If the same procedure is  repeated for the contribution of the anisotropy term of the Hamiltonian only, we find the equations: 
\begin{widetext}
\begin{eqnarray}
\frac{d\langle s_i\rangle}{dt} &=& \frac{\omega_a}{1+\lambda^2 }\,\epsilon_{ijk} \Big(  n_j n_l \langle s_k s_l \rangle - \lambda\, \epsilon_{jlm}\, n_l n_p \langle s_k s_m s_p \rangle \Big) -\frac{2D}{\left(1+\lambda^2  \right)^2}\, \langle s_i\rangle \label{Average_Aniso1} \\
\frac{d\langle s_i s_j \rangle}{dt} &=&\frac{\omega_a}{1+\lambda^2}\, \epsilon_{imn}\, \Big( n_m n_p\langle s_j s_n s_p\rangle - \lambda\, \epsilon_{mpq}\,  n_p n_r\langle s_j s_n s_r s_q\rangle \Big) + \frac{D}{\left( 1+\lambda^2\right)^2}\Big( \delta_{ij}\langle s_n s_n \rangle - 3\langle s_i s_j \rangle \Big) \nonumber\\
&&+\left(i \leftrightarrow j \right) \label{Average_Aniso2}
\end{eqnarray}
\end{widetext}
where $\omega_a\equiv 2K_a/\hbar$ is the effective field corresponding to the  anisotropy. The quadratic terms in  the RHS of eq.(\ref{sLLG}), when magnetic anisotropy is present, imply that eq.(\ref{Average_Aniso1}) depends  on three--point moments, and eq.(\ref{Average_Aniso2}) on four--point moments. And if we try to deduce the evolution equations for these moments, they will, in turn, depend on even higher moments.

Any treatment of these equations, therefore, involves closure assumptions, as we will discuss in the next section.

Let us,  now, consider, more than one spin, but with local interactions. For a collection of $N$ interacting spins we, apparently, have a straightforward generalization of the former expressions, the arguments of the probability $P(\left\{{\bm\sigma}^{I}\right\},t)$ just acquire indices, labeling the spins $1\le I\le N$, ${\bm s}^{I}(t)$ to be in a magnetic state ${\bm\sigma}^I$ at a given time $t$. However there's more to be said. 

For a given site $I$, the noise field $\bm{\eta}^I(t)$ is drawn from a known distribution $\rho(\bm{\eta}^I(t))$. The coupling with the spins leads to an induced distribution, $P(\left\{{\bm\sigma}^{I}\right\},t)$. If $\rho(\bm{\eta}^I(t))$ factorizes over the sites, $I$, i.e.
\begin{equation}
\label{rhoetafactor}
\rho(\{\bm{\eta}\})=\prod_{I=1}^N\,\rho(\bm{\eta}^I),
\end{equation}
then the factorization holds only for the measure of the noise. The expression for $P(\left\{{\bm\sigma}^{I}\right\},t)$ takes the form 
\begin{widetext}
\begin{equation}
\label{ProbN1}
P(\left\{{\bm\sigma}^{I}\right\},t) = \int\,\left(\prod_{I=1}^N \left[ {\mathscr D}{\bm\eta}^I(t) \right]\,\rho(\bm{\eta}^I)\right)
\prod_{K=1}^N\,{\bm\delta}\left( \bm{s}^K(\{\bm{\eta}\})-\bm{\sigma}^K\right).
\end{equation}
\end{widetext}

The same reasoning as before conducts to a formal master equation for $P$ as
\begin{equation}
\frac{\partial P(\left\{{\bm\sigma}^{K}\right\},t)}{\partial t}=-\frac{\partial J^I_i(\left\{{\bm\sigma}^K\right\},t)}{\partial\sigma^I_i},
\end{equation}
where the sum on $I$ runs from $1$ to $N$ as a repeated index. The evolution equation for any, equal--time, correlation function of the spin variables can then be expressed as
\begin{widetext}
\begin{eqnarray}
\frac{d}{dt} \langle s^{I_1}_{i_1} (t) \cdots s^{I_K}_{i_k}(t) \rangle &=& \int \prod_{M=1}^Nd\bm{\sigma}^M\, J^L_{j}\left(\left\{{\bm\sigma}^P\right\},t \right) \frac{\partial}{\partial \sigma^L_j} \left(\sigma^{I_1}_{i_1}\cdots \sigma^{I_K}_{i_k} \right).
\end{eqnarray} 
\end{widetext}
An explicit expression for  the probability flow $J^I_i$ is, in general, very challenging to find. This, of course, does not imply that the spins do not interact--indeed, if we attempt to resolve the $\delta-$functional constraint and obtain a functional integral over the $\bm{\sigma}^I$, we shall not, necessarily, find that it factorizes over the sites. However, if $\bm{s}^I(\{\bm{\eta}\})=\bm{s}^I(\bm{\eta}^I)$, i.e. that the spin at site $I$ depends only on the realization of the noise on the same site, then the measure over the spins will factorize as well,
\begin{equation}
\label{MFA}
P(\{\bm{\sigma}\},t) = \prod_{I=1}^N\,P(\bm{\sigma}^I,t)
\end{equation}
and the mean field approximation will be exact. 
 
The exchange interaction, that  controls the local alignment and order of spins is  defined by the following expression \cite{skubic2008method,evans2014atomistic}:
\begin{equation}
{\mathscr A}_{ex}=- \sum_{I,J\neq I}^N {\sf J}_{IJ}\, {\bm s}^I(t) . {\bm s}^J(t)  \label{Ex_Hamiltonian}
\end{equation}
where $\bm{s}^I(t)$ and $\bm{s}^J(t)$ are the values of neighboring spins at time $t$, and ${\sf J}_{IJ}$ is the strength  of the exchange interaction between these spins. This expression, indeed, appears in the sLLG and can be identified precisely with the exchange Hamiltonian at equilibrium. 

When working out of equilibrium, the mean--field approximation to the dynamics, described in the $P(\bm{\sigma}^I,t)$, by a two--spin interaction,  is reduced  by an averaging method \cite{anderson1953exchange,reimers1991mean} to that of one spin in an effective field. The exchange interaction is then described by 
${\mathscr A}=-{\sf J}_{ex}s_i(t)\cdot\langle s_i(t)\rangle$, where ${\sf J}_{ex}=n_v {\sf J}_{IJ}$, where  $n_v$ is  the number of  neighboring spins for any spin in its first and second shells of neighbors.

According to appendix \ref{App1}, the contribution of the exchange interaction to the moment equations can now be computed by noting that 
\begin{eqnarray}
&&\int\,[{\mathscr D}\bm{\eta}]\,\rho(\bm{\eta}){\bm\delta}(\bm{s}[\bm{\eta}]-\bm{\sigma})A_i(\bm{s}(t),\left\langle\bm{s}\right\rangle(t))\nonumber\\
&=& A_i(\bm{\sigma},\left\langle\bm{s}\right\rangle(t)) \int\,[{\mathscr D}\bm{\eta}]\,\rho(\bm{\eta}){\bm\delta}(\bm{s}[\bm{\eta}]-\bm{\sigma})
\label{extracting-exch}
\end{eqnarray}
since 
\begin{equation}
\label{avespin}
\left\langle \bm{s}\right\rangle(t) = \int\,d\bm{\sigma}\,\bm{\sigma}\,P(\bm{\sigma}=\bm{s}(t),t)
\end{equation}
depends only on time and the path integral does not, since it's time translation invariant. Because each spin $I$ has now the same effective field as any othe spin, the index $I$ can be safely dropped. Then the exchange interaction contributes to the moment equations as~:
\begin{widetext}
\begin{eqnarray}
\frac{d\langle s_i\rangle}{dt} &=& \frac{\lambda\omega_{ex}}{1+\lambda^2}\Big(\langle s_i\rangle\langle s_k s_k\rangle-\langle s_k\rangle\langle s_ks_i\rangle\Big)-\frac{2D}{\left(1+\lambda^2  \right)^2}\, \langle s_i\rangle,  \label{Average_Exchange1} \\
\frac{d\langle s_i s_j \rangle}{dt} &=& \frac{\omega_{ex}}{1+\lambda^2}\epsilon_{ikl} \Big( \langle s_k\rangle \, \langle s_j s_l\rangle -\lambda\, \epsilon_{kmn}\, \langle s_m\rangle\, \langle s_j s_l s_n\rangle \Big)  +\frac{D}{\left( 1+\lambda^2\right)^2} \Big( \delta_{ij}\langle s_n s_n \rangle - 3\langle s_i s_j \rangle \Big) \nonumber\\
&&+\left(i \leftrightarrow j \right). \label{Average_Exchange2}
\end{eqnarray}
\end{widetext}
with $\omega_{ex}=2{\sf J}_{ex}/\hbar$ is the exchange pulsation. Simplifications have been performed in eq.(\ref{Average_Exchange1}), because $\epsilon_{ijk}\langle s_j\rangle\langle s_k\rangle=0$ in the mean-field approximation. 

Now if a classical ferromagnet in an anisotropic and external fields is considered, we have to compute the contribution of each interaction to the moment equations, and their final form can be obtained by straightforwardly adding the RHSs. The only subtle point is, of course, that the longitudinal damping contribution shouldn't be over-counted. The full expressions aren't very illuminating as such; suffice to stress that they have been obtained under very few and tightly controlled assumptions. These couple the moments of different orders in an open hierarchy, that can't be easily solved, however (as is the case for the Gaussian distribution, for instance). Therefore we shall construct a framework, where ways to close the hierarchy can be tested  in a numerically useful manner. 

For Gaussian distributions of the noise, Wick's theorem allows us to obtain all the moments in terms of the first- and second-order moments only~\cite{justin1989quantum} and, therefore, close the systems of equations. We thus assume that Eqs.(\ref{Average_Zeeman1},\ref{Average_Zeeman2},\ref{Average_Aniso1},\ref{Average_Aniso2},\ref{Average_Exchange1}) and (\ref{Average_Exchange2}) give enough information for the simulation of the average magnetization dynamics, using the Gaussian closure.

However, we would like to check whether the distributions of our spin variables might deviate, in general, from a Gaussian distribution and how it might be possible to explore the validity of non-Gaussian closure schemes. How to close this hierarchy in such a fashion  
will be the subject of the following section.

\section{Closing the hierarchy}\label{closure}
In the previous sections, under controlled assumptions, equations governing the dynamics of all the moments have been derived and explicitly given for the first and second order moments of the spin variables. These averaging techniques give rise to an open hierarchy (possibly infinite if all the moments are required) of equations for the moments--that is not closed. In order to solve such a system, and to deduce the consequences for the magnetization dynamics itself (i.e. the first moment), this hierarchy must be closed in some way. 

We use closure methods, inspired from turbulence theory \cite{frisch1995turbulence,mellor1974hierarchy} and dynamical systems \cite{nicolis2012dynamics} and carry out numerical tests, in order to assess their range of validity. In order to check the consistency of these assumptions, on which the closure methods are founded, with respect to the model at hand, a reference model is required.

\subsection{Reference model}\label{Section_reference_model}
Atomistic spin dynamics (ASD) simulations are the usual way to solve the sLLG equation (\ref{sLLG}), with a white--noise process shown by eq.(\ref{White_noise}). Using this equation for ASD simulations was justified in refs.~\cite{antropov1995ab,antropov1996spin,skubic2008method}, and since, several numerical implementations have been reported \cite{nowak2005spin,beaujouan2012anisotropic,evans2014atomistic}, including the exchange interaction, the treatment of external and anisotropic magnetic fields and temperature. 

With identical sets of initial conditions, many configurations of $N$ spins are generated and for each individual spin, an sLLG equation (eq.~(\ref{sLLG})) is integrated. These integrations are performed by a third-order Omelyan algorithm, which preserves the symplectic properties of the sLLG equation \cite{krech1998fast,omelyan2003symplectic,ma2008large}. More details of this integration method are provided in previous works \cite{beaujouan2012anisotropic,beaujouan2012thermal}. These ASD simulations are performed for different noise realizations and averages are taken. 

 In practice we find that it is possible to generate a sufficient number of  noise configurations, so that the map induced by the stochastic equations, as  the result of this averaging procedure, realizes the exact statistical average over the noise\cite{mendez2011instabilities,mendez2014role}. These averages define, therefore, our reference model. 

\begin{figure}[thp]
\centering
\resizebox{0.99\columnwidth}{!}{\includegraphics{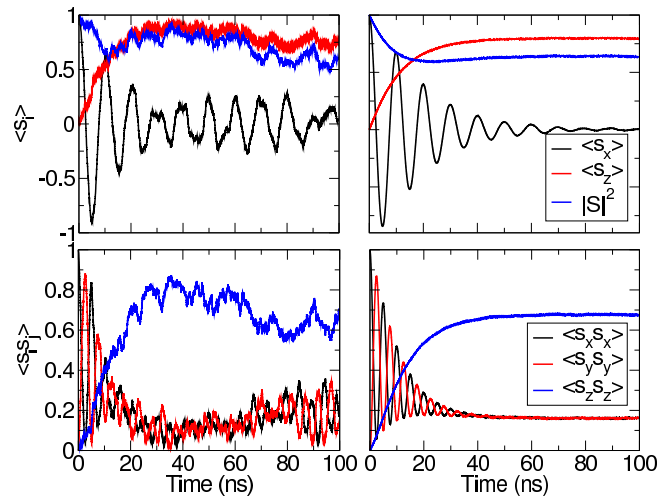}}
\caption{
Random magnetization dynamics of paramagnetic spins in a constant magnetic field. The upper graphs plot some of the first--order moments and the norm of the averaged magnetization, when the lower graphs plot the diagonal elements only of the matrix of the second--order moments. The averages over 10 paramagnetic spins only are shown on the left, whereas $10^4$ spins are shown on the right. Parameters of the simulations~: \{$D=5.10^{-2}$~rad.GHz; $\lambda=0.1$; $\vec{\omega}=(0,0,0.63)$~rad.GHz; timestep $\Delta t=10^{-3}$ ns\}. Initial conditions: $\vec{s}(0)=\left(1,0,0 \right)$, $\langle s_i(0) s_j(0) \rangle=0$ exept $\langle s_x(0)s_x(0) \rangle=1$. \label{Reference_model}}
\end{figure}

Figure~\ref{Reference_model} presents how effective this averaging procedure can be. The example of convergence toward statistical average for paramagnetic spins is shown.   
From Fig.\ref{Reference_model} one readily grasps  that increasing the number of spins (or, equivalently, realizations) does accelerate convergence toward the true averaged dynamics and $10^4$, interacting or not, spins can be taken  to be enough for practical purposes. This fixes statistical errors to sufficiently low level to draw accurately the desired average quantities, that can be lowered consistently by increasing the number of spins in ASD simulations if necessary.  

From now on, this averaging procedure is used in order to check the consistency of the closure assumptions presented in below.

\subsection{Gaussian Closure Assumption}\label{Gclosure}

The simplest possible way to close the hierarchy, that's consistent with Gaussian statistics, is to assume that the vacuum state is known, namely, that the second order cumulant goes to a given matrix $\chi$, i.e. $\langle s_i s_j\rangle-\langle s_i\rangle \langle s_j\rangle=\chi_{ij}$. This approximation has been studied by Ma and Dudarev~\cite{ma2011langevin}, where the matrix $\chi$ vanishes  for an ideal paramagnet. In a constant precession field, eq.(\ref{Average_Zeeman1}) becomes
\begin{eqnarray}
\frac{d\langle s_i\rangle}{dt} &=& \frac{1}{1+\lambda^2}\epsilon_{ijk} \Big(\omega_j \,\langle s_k\rangle-\lambda \,\epsilon_{jlm}  \omega_l \langle s_k\rangle\langle s_m \rangle \Big)\nonumber\\
&&+\frac{\lambda}{1+\lambda^2}\left(\omega_i{{\sf{Trace}}(\chi)}-\chi_{ij}\omega_j\right)\nonumber\\
&&-\frac{2D}{\left(1+\lambda^2  \right)^2}\, \langle s_i\rangle,
\label{SystemZeeman_closure1}
\end{eqnarray}
and presents some advantages and drawbacks. Let us define the vector $\bm{b}$ by the expression
\begin{equation} 
\label{vectorb}
b_i\equiv\left(\omega_i{{\sf{Trace}}(\chi)}-\chi_{ij}\omega_j\right).
\end{equation}
At equilibrium, the RHS of eq.(\ref{SystemZeeman_closure1}) vanishes. This provides an equation for the equilibrium value of the magnetization, $\langle s_i\rangle^{\sf{eq}}$, that is  proportional  to the precession field $\omega_i$ and a relation between the vector $\bm{b}$ and this equilibrium value:
\begin{equation}
\label{vectorbmageq}
b_i=\frac{2D}{\lambda(1+\lambda^2)}\langle s_i\rangle^{\sf{eq}}.
\end{equation}
This means that the value of the magnetization at equilibrium, $\langle s_i\rangle^{\sf eq}$, remains to be determined. 
This doesn't make this model very predictive, and constitutes a first, intrinsic, drawback.
Replacing  eq.~(\ref{vectorbmageq}) in eq.~(\ref{SystemZeeman_closure1}) leads to  
\begin{eqnarray}
\frac{d\langle s_i\rangle}{dt} &=& \frac{1}{1+\lambda^2}\epsilon_{ijk} \Big(\omega_j \,\langle s_k\rangle-\lambda \,\epsilon_{jlm}  \omega_l \langle s_k\rangle\langle s_m \rangle \Big)\nonumber\\
&&+\frac{2D}{\left(1+\lambda^2  \right)^2}\, \left(\langle s_i\rangle^{\sf eq}-\langle s_i\rangle\right),
\label{Bloch1}
\end{eqnarray}
which can be considered a generalization of Bloch's equation, that  includes a transverse damping. It has  many features in common with  the Landau-Lifshitz-Bloch equation derived by Garanin \cite{garanin1997fokker}. Given $\bm{b}$ or equivalently $\langle s_i\rangle^{\sf eq}$, eq.(\ref{Bloch1}) is straightforward to solve. 

When $D=0$, i.e. thermal effects can be neglected, damping does not affect the longitudinal part of the magnetization, which is, also, captured by the LLG equation.  However, the meaning of a statistical averaging procedure when $D=0$ can be questioned. Indeed, the set of equations (\ref{Average_Zeeman1}) and (\ref{Average_Zeeman2}) was derived using the value of $P(\bm{\sigma},t)$. This probability density is given by a functional integral over the noise realizations, and, of course, in that case, becomes a projector on a single configuration, since it collapses to a $\delta-$functional. This might be consistent, if such an equilibrium configuration is, indeed, unique.

Moreover, if $\omega_i$ is a mean-field exchange term only, no precession around this field occurs and eq.(\ref{Bloch1}) is purely longitudinal. For a ferromagnet, this has the consequence that it is, then impossible to capture, in this way,  any dynamics that would appear through   the frequency of the exchange constant.   

At this point, one understands that other closure methods might be considered, assuming Gaussian dynamics, i.e. that non--quadratic cumulants vanish and, nonetheless, consistent with  the interactions we want to consider. Once given the order of mixed averaged equations, this assumption leads to direct relations between third-- and fourth--order moments, and lower-order moments, known as Wick's theorem\cite{zinn2007phase}. This approach, called the Gaussian Closure Assumption (GCA) in this context, has been explored briefly in previous works \cite{tranchida_closing_2016,thibaudeau2015non}.

Denoting the cumulant of any stochastic spin vector variable ${\bm s}$ by double brackets $\langle \langle.\rangle\rangle$ \cite{van1992stochastic}, one has~:
\begin{eqnarray}
\langle\!\langle s_is_js_l\rangle\!\rangle &=&\langle s_is_js_l\rangle-\langle s_is_j\rangle\langle s_l\rangle-\langle s_is_l\rangle\langle s_j\rangle\nonumber\\
&-&\langle s_js_l\rangle\langle s_i\rangle+2\langle s_i\rangle\langle s_j\rangle\langle s_l\rangle\label{Cumulants3}
\end{eqnarray}
for any combination of the space indices for the third-order cumulant and
\begin{widetext}
\begin{eqnarray}
\langle\!\langle s_is_js_ls_m\rangle\!\rangle &=&\langle s_is_js_ls_m\rangle-6\langle s_i\rangle\langle s_j\rangle\langle s_l\rangle\langle s_m\rangle-\langle s_is_j\rangle\langle s_ls_m\rangle\nonumber\\
&&-\langle s_is_l\rangle\langle s_js_m\rangle-\langle s_is_m\rangle\langle s_js_l\rangle-\langle s_i\rangle\langle s_js_ls_m\rangle\nonumber\\
&&-\langle s_j\rangle\langle s_is_ls_m\rangle-\langle s_l\rangle\langle s_is_js_m\rangle-\langle s_m\rangle\langle s_is_js_l\rangle\nonumber\\
&&+2\left\{\langle s_i\rangle\langle s_j\rangle\langle s_ls_m\rangle+\langle s_i\rangle\langle s_l\rangle\langle s_js_m\rangle+\langle s_i\rangle\langle s_m\rangle\langle s_js_l\rangle\right.\nonumber\\
&&+\left.\langle s_j\rangle\langle s_l\rangle\langle s_is_m\rangle+\langle s_j\rangle\langle s_m\rangle\langle s_is_l\rangle+\langle s_l\rangle\langle s_m\rangle\langle s_is_j\rangle\right\}\label{Cumulants4}
\end{eqnarray}
\end{widetext}
for any combination of the space indices for the fourth-order cumulant. GCA implies that, for every time $t$, $\langle\!\langle s_is_js_k\rangle\!\rangle=0$ and $\langle\!\langle s_is_js_ks_l\rangle\!\rangle=0$. Thus the following relationships apply~:
\begin{eqnarray}
\langle s_i s_j s_k\rangle                         &=& \langle s_i\rangle\langle s_j s_k\rangle + \langle s_j\rangle\langle s_i s_k\rangle \nonumber\\
                                                   &~& + \langle s_k\rangle \langle s_i s_j\rangle -2\langle s_i\rangle\langle s_j\rangle\langle s_k\rangle \label{Gaussian_closure1}, \\
\langle s_i s_j s_k s_l\rangle                     &=&\langle s_i s_j\rangle \langle s_k s_l\rangle+ \langle s_i s_k\rangle \langle s_j s_l\rangle \nonumber  \\
                                                   &~&+ \langle s_i s_l\rangle \langle s_j s_k\rangle -2 \langle s_i\rangle \langle s_j\rangle \langle s_k\rangle \langle s_l\rangle, \label{Gaussian_closure2}
\label{KY}
\end{eqnarray}
relating thereby the third and fourth moments with the first and second ones only. Equations (\ref{Gaussian_closure1}) and (\ref{Gaussian_closure2}) have to be injected into Eqs.(\ref{Average_Zeeman1},\ref{Average_Zeeman2},\ref{Average_Aniso1},\ref{Average_Aniso2},\ref{Average_Exchange1}) and (\ref{Average_Exchange2}) respectively. Because of the form these equations assume, they were called dynamical Landau-Lifshitz-Bloch (d-LLB) equations\cite{tranchida_closing_2016,thibaudeau2015non} and reveal, for both the first and second moments, a longitudinal contribution, proportional to the amplitude of the noise,  to the damping of  the average magnetization.

\begin{figure}[thp]
\centering
\resizebox{0.99\columnwidth}{!}{\includegraphics{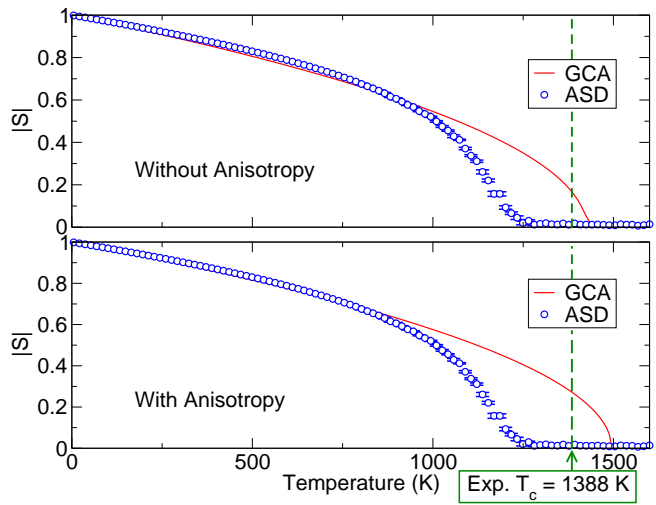}}
\caption{Equilibrium magnetization norm vs temperature for hcp-Co, without (on top), and with (below) the anisotropic interaction. The solid line plot the result of the GCA applied on the third-order moments. Open circles (with error bars) plot the ASD results, performed with the s-LLG equation. The experimental Curie temperature $T_c$ for hcp-Co is also reported.
\label{Gaussian_comparison_T}
}
\end{figure}

\begin{figure}[thp]
\centering
\resizebox{0.99\columnwidth}{!}{\includegraphics{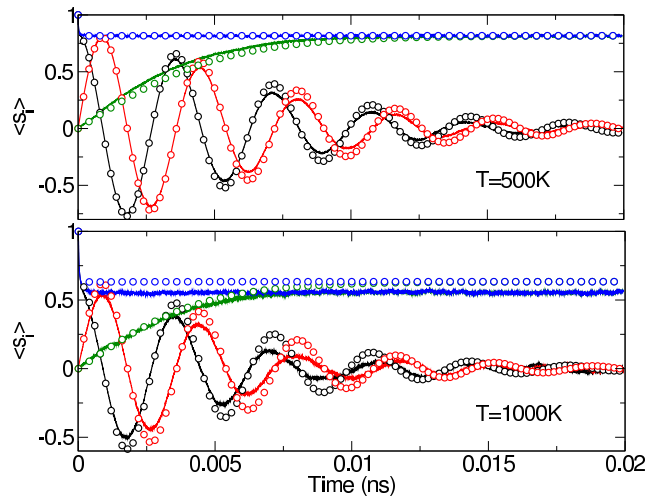}}
\caption{
Sub-figures show the relaxation of the average dynamics up to 20ps for $500$K and $1000$K, under a constant external magnetic induction of 10 T, applied on the $z$-axis for initial conditions $s_y(0)=1$ on each spin, and $\lambda=0.1$. ASD are in solid lines ($\langle s_x\rangle$ in black, $\langle s_y\rangle$ in red, $\langle s_z\rangle$ in blue and $|s|$ in green), whereas d-LLB with GCA are in open circles (see text). 
\label{Gaussian_comparison_M}}
\end{figure}

\begin{figure}[thp]
\centering
\resizebox{0.99\columnwidth}{!}{\includegraphics{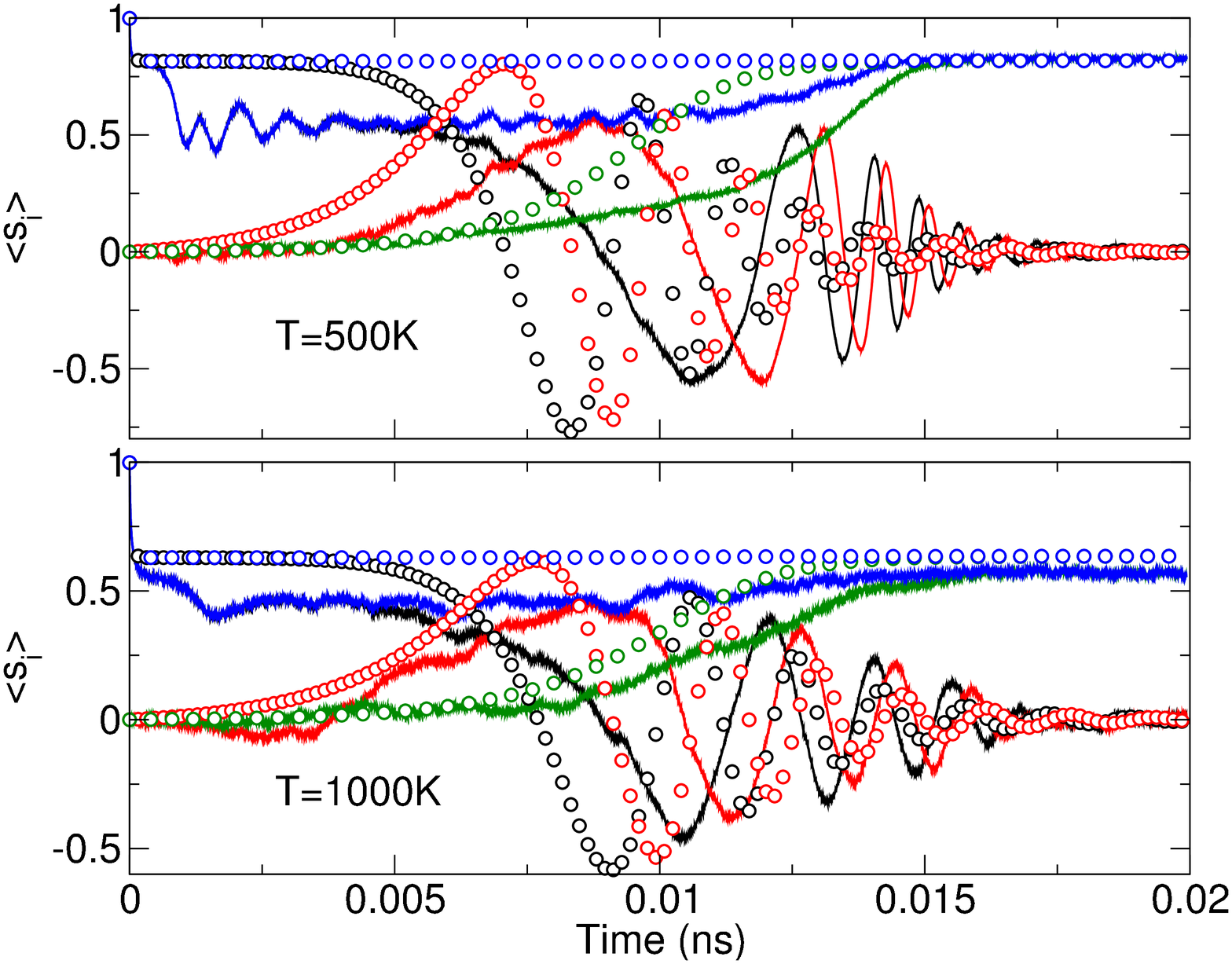}}
\caption{
Sub-figures show the relaxation of the average dynamics up to 20ps for $500$K and $1000$K, under an uniaxial anisotropic field (K$_a$=4.17 meV) and a constant external magnetic induction of $0.1$ T,  both applied along the $z$-axis. The initial conditions are $s_x(0)=1$ on each spin, and $\lambda=0.1$. ASD are depicted in solid lines ($\langle s_x\rangle$ in black, $\langle s_y\rangle$ in red, $\langle s_z\rangle$ in blue and $|s|$ in green), whereas d-LLB with GCA are shown in open circles (see text). 
\label{Gaussian_comparison_Ka}}
\end{figure}

Simulations of hcp-Co have been performed and are depicted in figures \ref{Gaussian_comparison_T}, \ref{Gaussian_comparison_M} and \ref{Gaussian_comparison_Ka}. These figures compare the GCA, applied to the third moments according to eq.(\ref{Gaussian_closure1}), with  the ASD calculations,  for an hexagonal $22\times 22\times 22$-supercell. The first and second nearest neighbor shell are taken into account  for the exchange interaction, and its value, taken from references\cite{pajda2001ab,lounis2010mapping}, is ${\sf J}_{IJ}=29.79$ meV for each atomic bond of the first nearest neighbors, and ${\sf J}_{IJ}=3.572$ meV for the second nearest neighbors. 
The anisotropy energy for hcp-Co, is given to $K_a = 4.17.10^{-2}$ meV for each spin, also according to references\cite{pajda2001ab,lounis2010mapping}. 

The magnetic analog  of the Einstein relation can be  introduced in order to relate the amplitude $D$  of the noise to the temperature of the bath \cite{neel1953thermoremanent,brown1979thermal}:
\begin{equation}
D=\frac{\lambda k_B T}{\hbar \left(1+\lambda^2\right)}
 \label{Einstein_relation}
\end{equation}
The conditions for the validity  of such an expression aren't immediately obvious (especially, in our case, the equilibrium condition that is necessary for the derivation of a fluctuation--dissipation relation).
However, this discussion is beyond the scope of this work, and eq.~(\ref{Einstein_relation}) is assumed to be valid. This allows us to replace averages over the noise  by corresponding thermal averages. The reason this is useful is that, in practice, one is measuring thermal averages and is interested in the Curie point. 
 
Figure~\ref{Gaussian_comparison_T} plots the average magnetization norm versus the temperature for hcp-Co with and without the anisotropic contribution, over a long simulation time, assuming the system at equilibrium. The GCA on the third-order moments matches reasonably well the ASD calculations--and without requiring  prior knowledge of the equilibrium magnetization value. Thus, the GCA can be considered to be valid at least up to half  the Curie temperature, $T_c/2$. For higher temperatures, however, a significant departure from the ASD calculations is observed. This is not surprising because the correlation length of the connected real-space two-point correlation function at equilibrium grows without limit when $T$ approaches $T_c$. Magnetization fluctuations occur  in blocks of all sizes up to the size of the correlation length, but fluctuations that are significantly larger are exceedingly rare.  Interestingly, within the GCA, the equilibrium magnetization passes through  a critical transition, from a ferromagnetic to a paramagnetic phase, driven by the  temperature.

Figure~\ref{Gaussian_comparison_M} plots the time dependence  of the average magnetization for hcp-Co for  an external magnetic field of $10$T along the $z$-axis, without any internal anisotropic contribution. The value of the external magnetic field is conveniently chosen to hasten the convergence of large ASD simulations. Besides, as the closure assumption does not rely on the intensity of the Zeeman interaction, any value of the field can be used. For T$=500$K, the GCA appears to be a good approximation and the two models are in good agreement. For T$=1000$K, the validity of the GCA becomes more questionable, and the two models present now some marked differences, in particular regarding  the norm of the average magnetization and $\langle s_z\rangle$ at equilibrium, less so  in the transient regime. 

 Another interesting feature of this figure is the presence of two regimes for the magnetization dynamics. The first one is an extremely short thermalization regime. Because the exchange pulsation is the fastest pulsation in the system, the magnetization norm sharply decreases in order to balance the exchange energy with the thermal agitation. The second regime is the relaxation around the Zeeman field itself. The GCA model and the ASD simulations are in good agreement concerning the characteristic times of both these regimes.

Figure~\ref{Gaussian_comparison_Ka} displays the non-equilibrium profile  of the average magnetization for hcp-Co assuming  uniaxial anisotropy, oriented along the $z$-axis, along with a small Zeeman field, also along the $z$-axis, which ensures that the average magnetization aligns itself along the $+z$ direction. For T$=500$K, the GCA leads to  the same equilibrium magnetization as the ASD, whereas for T$=1000$K, the average magnetization norm, and the average magnetization along the $z$-axis, calculated by the GCA, show deviations  from the ASD calculations. Moreover, for the temperatures used, the GCA, also,  fails to match the transient dynamics of the relaxation. The ASD calculations indicate a lag for the magnetization, compared to the results obtained by the GCA and even if the precession frequency of the two models is the same, their dynamics are correspondingly  shifted.

However, we saw that the GCA does predict an  equilibrium magnetization value consistent with that  of the ASD simulations, for Zeeman, exchange and anisotropic energies, up to $T_c/2$. In a constant field, below this temperature, the transient regimes also correctly match those obtained from  the ASD. For temperatures higher than $T_c/2$, departures from the ASD simulations are observed, both for the equilibrium magnetization values and for the details of the transient regimes.

The GCA is, indeed, not suited for describing magnon interactions, that involve more than two magnons, since Wick's theorem implies that all such processes factorize. 

The reason isn't the  validity of the closure assumption itself, but that interacting magnon modes are generated inside the large $22^3$ ASD cell. Indeed, with a local anisotropy field only, the energetics of the spins is less constrained because individual spins may equilibrate along or in opposite direction of the anisotropy-axis. For a large but finite ASD cell, with periodic boundary conditions, this has as consequence to generate local spin configurations (small "sub-cells" inside the large cell) due to the different realizations of the noise. In order to dissipate these sub-cells,  additional, internal, magnons are produced (and reflected by boundaries);  and their collective motion  cannot be described by average thermal modes only, as we can see in  the very beginning of the transient regimes of both graphs of Fig.\ref{Gaussian_comparison_Ka}. The GCA model, that simulates the average over the repetitions of one single spin, is unable to recover these extra magnon modes, corresponding to spin waves generated by sub groups inside the large ASD cell. As a consequence, the effective precession around the anisotropy field is shifted and delayed, in  the ASD simulation.

In fig.~\ref{Gaussian_comparison_M}, only the Zeeman and the exchange interactions are considered. Thus, each individual spin has only one possible equilibrium position, and the property of ergodicity is preserved. However, as can be seen from fig.~\ref{Gaussian_comparison_Ka}, when  uniaxial anisotropy is added to the two former interactions, each individual spin has now two equilibrium positions. Even if, due to the presence of the Zeeman interaction, these two equilibrium positions are not equally probable, they both have a non-zero probability to occur. Therefore, at each realization of the ASD anisotropic simulation, different local spin configurations are occurring. This leads to different,  transient, values for the moments, that depend strongly  on the noise realizations, and, thus, to departures from ergodicity.

In order to enhance the agreement for the equilibrium magnetization state of both ASD and averaged models, another closure method, more sophisticated than the GCA, will be considered in the following.

\subsection{Non-Gaussian closure}\label{NGclosure}
This  method is  inspired by studies in chaotic dynamical systems, where elaborate moment hierarchies are typically encountered \cite{Levermore:1996ys,Eu:1998rc}. 

Closure relations can, indeed,  be derived for the hierarchy of moments for the invariant measure of dynamical systems~\cite{bobryk2011closure}. The proof relies on properties of the Fokker--Planck equation, and on the assumption of ergodicity \cite{nicolis1998closing}.

However, we saw in the previous section that, depending on the magnetic interactions that are at stake, departures from ergodicity can be observed in the ASD simulations of large cells. 

Therefore, since the Non-Gaussian Closure Assumption (NGCA) presented below is only expected to hold for ergodic situations, only the exchange and the Zeeman interactions will be  considered, or cases  when the Zeeman interaction is stronger than the anisotropic interaction, forcing each individual spin toward one possible equilibrium position.

The formalism can be presented as follows:
Assuming ergodicity and with the cumulant notation at hand, such a NGCA relation can be parametrized for a stochastic variable $\bm{s}$ as
\begin{eqnarray}
 \langle\! \langle s_i  s_j s_k \rangle\! \rangle &=& a^{(1)}_i \langle\! \langle s_j s_k \rangle\! \rangle + a^{(1)}_j \langle\! \langle s_i s_k \rangle\! \rangle + a^{(1)}_k \langle\! \langle s_i s_j \rangle\! \rangle \nonumber \\
                                                ~&~&+a^{(2)}_{ij} \langle s_k \rangle + a^{(2)}_{ik} \langle s_j \rangle + a^{(2)}_{jk} \langle s_i \rangle \label{Linear_Closure_Cumulants}
\end{eqnarray}
Coefficients $a^{(1)}_i$ and $a^{(2)}_{ij}$ are assumed not to depend on time, but only on system parameters, such as $D$, $\bm{\omega}$ and $\lambda$. These coefficients are assumed to be exactly zero when $D=0$, hence matching the GCA. These Non-Gaussian Closure Approximations (NGCA) are tested with eqs. (\ref{Average_Zeeman2}). The next logical step is to determine the values of these coefficients. As the third-order cumulants are symmetric under  permutation of the coordinate indices, the coefficients $a^{(2)}_{ij}$ are symmetric, too, and only nine coefficients are required, in all. 

According to Nicolis and Nicolis~\cite{nicolis1998closing}, it was stressed that these coefficients satisfy constraining identities, that express physical properties of the spin systems considered. However, finding the corresponding identities, in general, is quite non--trivial, and to circumvent this difficulty, a fully computational approach was chosen. It is useful to stress that this approach is not without its proper theoretical basis: these identities, indeed, express properties of the functional integral\cite{zinn2007phase}. 

ASD calculations are used to fit the coefficients $a^{(1)}_i$ and $a^{(2)}_{ij}$, for a given set of system parameters. At  a given time, a distance function $d$, defined from the results of ASD simulations and the new, closed model as
\begin{eqnarray}
d^2(t)&\equiv&\sum_{i=1}^3\left(\langle s_i(t)\rangle^{\sf{ASD}}-\langle s_i(t)\rangle\right)^2 \nonumber\\
              &~&+\sum_{i,j=1}^3\left(\langle s_is_j\rangle^{\sf{ASD}}(t)-\langle s_is_j\rangle(t)\right)^2
\label{Distance_function}
\end{eqnarray} 
is computed and a least-square fitting method is applied. In this distance expression, each term is  weighted equally to avoid any bias. At each step of the solver, a solution of the system of equations (eqs.~(\ref{Average_Zeeman1},\ref{Average_Zeeman2},\ref{Average_Aniso1},\ref{Average_Aniso2},\ref{Average_Exchange1}) and (\ref{Average_Exchange2}), closed by eq.~(\ref{Linear_Closure_Cumulants})) is computed, and the distance function is evaluated. From the evolution of this distance, the method determines a new guess for the coefficients $a^{(1)}_i$ and $a^{(2)}_{ij}$. When the distance reaches a minimum, the hierarchy is assumed to be closed with the corresponding coefficients.
 
In order to check the validity of the NGCA, this was applied for the simulation performed at T=$1000$K presented in the previous section because, in these situations, neither the equilibrium nor the transient regimes of the ASD simulations were recovered by the GCA. 

A new trial  is carried out  by performing again the simulation of the second part of fig.~\ref{Gaussian_comparison_M}. At the equilibration time, a minimum distance is found by considering a restriction to the third values only, thus we find
$a^{(1)}_3=0.145$ and $a^{(2)}_{33}=0.145$. All the other coefficients are assumed to be zero. As expected, these dimensionless coefficients are small, demonstrating a slight departure of the GCA, which has to increase when the temperature increases. The uniqueness  of these coefficients is not obvious and may depend on the choice of the distance function and its corresponding weights.

\begin{figure}[thp]
\centering
\resizebox{0.99\columnwidth}{!}{\includegraphics{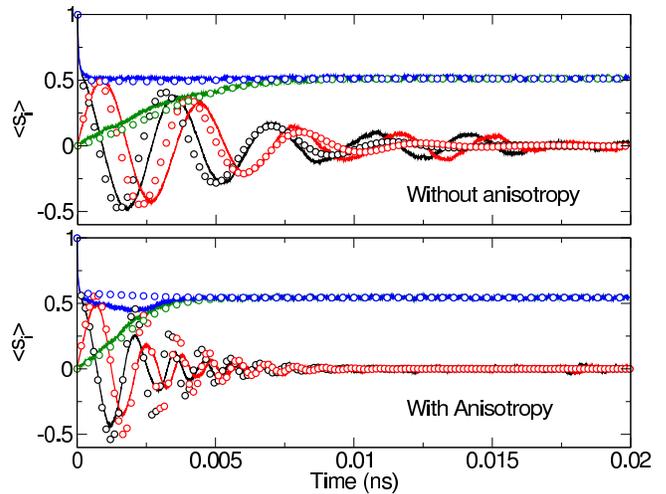}}
\caption{
Average dynamics up to 20ps for T=$1000$K, with $\lambda=0.1$ and initial conditions $\langle s_x(0) \rangle=1$. Above, the situation of a constant external field  of $10$T applied along the $z$-axis, and below the case of an uniaxial anisotropic field (K$_a=4.17$meV) and a constant external field of $10$T, both applied along the $z$-axis. ASD are in solid lines ($\langle s_x\rangle$ in black, $\langle s_y\rangle$ in red, $\langle s_z\rangle$ in blue and $|s|$ in green), whereas d-LLB with the NGCA approximation are in open circles.
\label{NGCA_1000K}
}
\end{figure}

Figure~\ref{NGCA_1000K} displays now the result of this closure, with and without anisotropic interaction. The two models present some slight differences in the beginning of the transient regime, but quickly match. This could be surely managed by increasing the number of distance points to match by relaxing all the coefficients. 

We now investigate the situation of including all interactions. For a different set of equations, a similar situation has already been investigated previously~\cite{thibaudeau2015non}, with a slightly different closure method. 
To close the hierarchy of moments in that case, an expression for the fourth--order moments $\langle s_i s_j s_k s_l\rangle $ is required. This is performed by assuming that the fourth-order cumulants are negligible (i.e. $\langle\!\langle s_is_js_ks_l\rangle\!\rangle=0$) and that each third-order moments are computed by eq.~(\ref{Linear_Closure_Cumulants}). Yet again, one can systematically improve on this hypothesis by increasing the number of desired coefficients up to this order such as
\begin{eqnarray}
\langle\!\langle s_is_js_ks_l\rangle\!\rangle&=&b^{(1)}_i\langle\!\langle s_js_ks_l\rangle\!\rangle + {\rm{perm.}}\nonumber\\
&&+b^{(2)}_{ij}\langle\!\langle s_ks_l\rangle\!\rangle + {\rm{perm.}}\nonumber\\
&&+b^{(3)}_{ijk}\langle s_l\rangle + {\rm{perm.}}\nonumber
\end{eqnarray} 
Once again, invariance under permutation of indices enhances  the symmetries of the $b^{(2)}$ and $b^{(3)}$ tensors and reduces the number of independent coefficients.

As a test, if we take  all these coefficients $b$ to be zero, at  equilibrium, a minimum is found with $a^{(1)}_3=0.07$ in the case where the $z$-axis is preferred. Fig.~\ref{NGCA_1000K} displays the results of the application of the NGCA in that case. Again, the equilibrium state is recovered, even if some differences remain  in the transient regime. 

By educated  guessing, we, thus,  saw that the NGCA allows to recover the equilibrium state of the magnetization, and a much better agreement between ASD and dLLB models is also observed during the transient regimes, than in the GCA.

\section{Conclusions}\label{conclusions}

The functional calculus has been applied to the study of the master equation for the probability distribution of magnetic systems, whether the contribution of individual magnetic moments can be resolved or not. The effects of the multiplicative, colored noise, whose physical origin is the fast stochastic fields that are relevant for current experiments, have been described under controlled analytical approximations and explicit expressions for the master equation have been deduced.

This  formalism  was applied  to  the dynamics of a system of coupled spins and used to equations for the evolution to equilibrium of certain  correlation  functions. In  the white--noise limit, the well-known Fokker--Planck equation was recovered, whereas in the case of the Ornstein-Uhlenbeck process, a new equation for the  probability density was derived. This equation explicitly displays  the correction terms, that appear in first order of the auto--correlation time expansion to  and the white-noise limit is, indeed,  recovered, when $\tau=0$. 

In the Markovian limit, the system of coupled equations for the  spin correlation functions was obtained and solved  for three fundamental magnetic interactions (Zeeman energy, exchange interaction, and uniaxial anisotropy). These equations give rise to infinite hierarchies of equations for the moments of the spin components, and two  methods were introduced, in order to close the hierarchy. In order to check the consistency of  these methods, results, obtained by numerical resolution,  were compared to stochastic simulations, performed using a completely independent, atomistic spin dynamics (ASD) formalism. 

When the magnetic interaction includes the exchange (in the mean-field approximation) and the Zeeman interactions only, the GCA proved sufficient to recover both, the transient regime and the equilibrium state, of the average magnetization, for a broad range of temperatures,  up to half  the Curie temperature for  ferromagnetic materials. When the anisotropy energy contribution is included, the probability flow on the space of spin configurations can become non--ergodic and the Gaussian approximation is expected to have problems. Indeed,  the GCA can describe  the equilibrium state for the same range of temperatures as before but, as  ergodicity is lost, the transient regime of the ASD simulation becomes  biased by a strong dependance on the noise realizations. Because rare local spin configurations are  generated by ASD simulations, the average set of equations of the dLLB model captures  the mean magnetization and its variance only. These features  were shown by direct inspection of the transient regimes, that allows to detect the  temporal shift, that is  represented by a more delayed variance memory kernel than the approximation could provide. 

For temperatures far from  half  the Curie point, as non--Gaussian fluctuations become more and more relevant, the GCA fails correspondingly to recover even  the equilibrium average magnetization and a  NGCA, inspired by work in dynamical systems,  was introduced.  The NGCA, also, relies on ergodicity, but it can provide a correspondingly better  match between the ``average''  models and the  stochastic calculations near the equilibrium. This was illustrated for the case of  the exchange interactions, treated in the mean-field approximation and Zeeman fields included, but deserves a more detailed study. However once parametrized properly, this is  a simple and reliable tool for closing the hierarchy of magnetic equations and it does recover both the equilibrium value of the magnetization in temperature and provides  a better picture of the dynamics in the transient regime--but its full range of validity remains to be explored. 

\begin{acknowledgments}
JT acknowledges financial support through a joint doctoral fellowship ``CEA - Région Centre'' under the grant agreement number 00086667, and would also like to thank C. Serpico for helpful comments about this work. 
\end{acknowledgments}


\appendix
\section{Master equation for $P(\bm{\sigma},t)$}\label{App1}
In this appendix, we review the salient results  of refs.~\cite{fox1986functional,fox1986uniform,ramirez1988higher}, which are the foundation of the functional calculus approach leading to  the master equation for the probability density, $P(\bm{\sigma},t)$. We have implicitly chosen the Stratonovich convention for  the stochastic calculus, and the derived expressions can be recast  to any other  prescription\cite{aron2014magnetization,moreno2015langevin}.

To simplify forthcoming expressions, it's useful to write the stochastic Landau-Lifshitz equation (\ref{sLLG}) in the form~:
\begin{equation}
\dot{s_i}=A_i(\bm{s}(t))+e_{ia}(\bm{s}(t))\eta_a(t),
\label{Langevin}
\end{equation}
where
\begin{eqnarray}
A_i(\bm{s}(t))   &=&\frac{1}{1+\lambda^2}\epsilon_{ijk}\, s_k \left(\omega_j - \lambda \epsilon_{jlm}\omega_l s_m \right) \label{LLGtoLangevin}, \\
e_{ia}(\bm{s}(t))&=&\frac{1}{1+\lambda^2}\epsilon_{iaj}\, s_j,\label{Vielbein}
\end{eqnarray}
with ${\bm\omega}$ a functional of $\bm{s}$. In eq.(\ref{Langevin}), since $e_{ia}(\bm{s}(t))$ depends on the spin variables, the noise is multiplicative. Geometrically this means that the manifold defined by the spin variables, $s_i(t)$, is curved. Its metric may be reconstructed from the vielbein, $e_{ia}(\bm{s}(t))$. Then, the magnetization explores ''islands'' on the surface of a sphere of constant radius. 

The Langevin equation provides the rule for realizing the change of variables from the noise, $\bm{\eta}(t)$, to the spin variables, $\bm{s}(t)$ and, therefore, leads to the definition of their probability density $P(\bm{\sigma},t)$, from the partition function for the spin variables. This latter may be defined, in terms of the partition function of the bath, as the average value of ${\bm\delta}(\bm{\sigma}-\bm{s}[\bm{\eta}(t)])$~: 
\begin{eqnarray}
P(\bm{\sigma},t) &=& \langle {\bm\delta}\left(\bm{\sigma}-\bm{s}[\bm{\eta}(t)]\right) \rangle \\
                  &=&  \int \left[ {\mathscr D}\bm{\eta} \right] \rho(\bm{\eta}) {\bm\delta}\left(\bm{\sigma}-\bm{s}[\bm{\eta}(t)]\right), \label{ProbaS}
\end{eqnarray}
where the integral in eq.(\ref{ProbaS}) is a path integral over all the noise realizations \cite{kleinert2009path}. 

The probability density for the noise process $\rho(\bm{\eta})$ is defined by its functional expression~:
\begin{equation}
\rho (\bm{\eta} )=Z^{-1}\,e^{ -\frac{1}{2}\int dt dt' \eta_b(t) G(t-t')\eta_b(t') }
 \label{Noise_proba}
\end{equation}

In eq.(\ref{Noise_proba}), the density $\rho (\bm{\eta})$ is normalized by the partition function $Z$ in eq.(\ref{bath_part_fun}) and $G(t-t')$  is the functional inverse of the 2--point correlation function $C(t-t')$, defined by the relation:
\begin{equation}
\int dt'\, G(t-t') C(t'-t'') = \delta(t-t'')
 \label{Inverse_noise_relation}
\end{equation}

The expression for $P(\bm{\sigma},t)$, in eq.(\ref{ProbaS}) is formal: the measure, $[{\mathscr D}\bm{\eta}(t)]$ needs to be defined, and the three dimensional ${\bm{\delta}}-$functional, also, so the purpose of the following calculations is to render the expression well--defined, by obtaining the evolution equations for its moments, from which it may be reconstructed.  We shall show how this program can be realized, without imposing any additional conditions on the spectral properties of the noise--at least for the master equation for $P(\bm{\sigma},t)$, which for colored, multiplicative noise, is not, in general of Fokker--Planck form and, thus, cannot be determined exclusively by imposing general coordinate invariance of the manifold, which the spin variables explore--which is what happens for white noise. 

By computing the time derivative of $P(\bm{\sigma},t)$
\begin{equation}
\frac{\partial P(\bm{\sigma},t)}{\partial t} = \int \left[ {\mathscr D}\bm{\eta} \right] \rho(\bm{\eta}) \frac{\partial}{\partial t}{\bm{\delta}}\left(\bm{\sigma}-\bm{s}[\bm{\eta}]\right),
 \label{ProbaP1}
\end{equation}
the chain rule and because the ${\bm\delta}$-functional is symmetric through the functional derivative, 
\begin{equation}
\frac{\delta}{\partial s_i(t)}{\bm\delta}\left(\bm{\sigma}-\bm{s}(t)\right) = -\frac{\partial}{\partial \sigma_i} {\bm\delta}\left(\bm{\sigma}-\bm{s}(t)\right),
\label{Symetry}
\end{equation}
one finds the following divergence~:
\begin{eqnarray}
\frac{\partial P(\bm{\sigma},t)}{\partial t} &=&-\frac{\partial}{\partial \sigma_i}\cdot\int \left[ {\mathscr D}\bm{\eta} \right] \rho(\bm{\eta}) {\bm\delta}\left(\bm{\sigma}-\bm{s}[\bm{\eta}]\right) \dot{s}_i(t) \label{ProbaP2}
\end{eqnarray}
When $\dot{s}_i(t)$ is replaced by the RHS of eq.(\ref{Langevin}), 
\begin{eqnarray}
\frac{\partial P(\bm{\sigma},t)}{\partial t} &=&-\frac{\partial}{\partial \sigma_i}\cdot\Big(  \int \left[ {\mathscr D}\bm{\eta} \right] \rho(\bm{\eta}) \bm{\delta}\left(\bm{\sigma}-\bm{s}[\bm{\eta}]\right)\nonumber \\
                                             ~&~&\left[ A_i(\bm{s}(t))+e_{ia}(\bm{s}(t))\eta_a(t)\right]\Big)\label{ProbaP3}
\end{eqnarray}
and the RHS of eq.(\ref{ProbaP3}) consists of two terms, each one having a different physical meaning. In the Langevin equation, $A_i(\bm{s}(t),\bm{\omega}(t))$, is often denoted as a drift term, and has a deterministic nature. Therefore, the first term of eq.(\ref{ProbaP3}) is called its drift part, and denoted by ${\mathcal Drift}$:
\begin{equation}
{\mathcal  Drift}\equiv -\frac{\partial}{\partial \sigma_i}\cdot\int \left[ {\mathscr D}\bm{\eta} \right] \rho(\bm{\eta}) {\bm\delta}\left(\bm{\sigma}-\bm{s}[\bm\eta]\right) A_i(\bm{s}(t),\bm{\omega}(t))
\label{Drift1}
\end{equation}
This term describes the interactions  of the spin system. In the mean--field approximation, which is valid, trivially, for the case of a single spin, considered here, it is possible to write the drift term in local form (containing a finite number of derivatives, only):
\begin{equation}
{\mathcal Drift}\equiv -\frac{\partial}{\partial \sigma_i}\cdot\left(A_i(\bm{\sigma},\bm{\omega}(t)) P(\bm{\sigma},t) \right),
\label{Drift2}
\end{equation}
which is possible by expanding the functional $A_i(\bm{s}(t),\bm{\omega}(t))$ around ${\bm\sigma}$ and by performing the integration over the noise.

The second term, of eq.(\ref{ProbaP3}), which is built up from $e_{ia}(\bm{s}[\bm{\eta}(t)])\eta_a(t)$, the noise term, would lead to the diffusion term, in the case of white, additive, noise; therefore, we shall call it  the diffusion term, here, as well,  and denote it  by ${\mathcal Diff}$, keeping in mind, however, that this is an abuse of language: 
 \begin{equation}
{\mathcal Diff}\equiv -\frac{\partial}{\partial \sigma_i}\cdot\int\left[ {\mathscr D}\bm{\eta} \right] \rho(\bm{\eta}) \bm{\delta}\left(\bm{\sigma}-\bm{s}[\bm{\eta}]\right) e_{ia}(\bm{s}(t))\eta_a(t)
\label{Diffusion1}
\end{equation}
This term strongly depends on the spectral properties of the noise (white or colored), and on whether the noise is additive ($e_{ia}=\delta_{ia}$) or multiplicative ($e_{ia}$ is a function of $\bm{s}(t)$).

We shall now perform on the diffusion term (\ref{Diffusion1}) a transformation, similar to that for the drift term,  that led to  eq.(\ref{Drift2}). This will highlight the spectral properties of the noise, that play a key role in distinguishing its effects from those of white, additive, noise. Expanding the vielbein $e_{ia}$ once again and performing the functional integration over the noise gives 
\begin{equation}
{\mathcal Diff}=-\frac{\partial}{\partial \sigma_i} e_{ia}(\bm{\sigma})\int \left[ {\mathscr D}\bm{\eta} \right] \rho(\bm{\eta}) \delta\left(\bm{\sigma}-\bm{s}[\bm{\eta}]\right) \eta_a(t)
\label{Diffusion2}
\end{equation}
The Gaussian integral (\ref{Noise_proba}), that defines the distribution function of the noise, is used to deduce the following formula~:
\begin{equation}
\frac{\delta \rho(\bm{\eta})}{\delta \eta_a(t)}=-\int dt'\, G(t-t') \eta_a(t') \rho(\bm{\eta})
\label{Distrib_deriv1} 
\end{equation}
where $\delta/\delta \eta_a(t)$ is the functional derivative \cite{justin1989quantum} by $\eta_a(t)$.  Applying the functional inverse (\ref{Inverse_noise_relation}), one has
\begin{equation}
\eta_a(t) \rho(\bm{\eta})=-\int dt'\, C(t-t') \frac{\delta \rho(\bm{\eta})}{\delta \eta_a(t')},
\label{Distrib_deriv2} 
\end{equation}
which leads to  the Furutsu-Novikov formula \cite{Furutsu:1963sf,Novikov:1965fp}, once integrated over all the realizations of the noise. Inserting into eq.(\ref{Diffusion2}), dropping total derivatives, and taking out of the integral term(s) that depend on $\bm{\sigma}$ only, one finds~:
\begin{eqnarray}
{\mathcal Diff}&=&\int dt'\, C(t-t')\frac{\partial}{\partial \sigma_i}e_{ia}(\bm{\sigma}) \nonumber \\
            ~&~&\int\left[{\mathscr D}\bm{\eta} \right]{\bm\delta}\left(\bm{\sigma}-\bm{s}[\bm{\eta}]\right)  \frac{\delta\rho(\bm{\eta})}{\delta \eta_a(t')}
 \label{Diffusion3}
\end{eqnarray}
Peforming a partial integration in the path integral of eq.(\ref{Diffusion3}) we end up with the expression~:
\begin{eqnarray}
{\mathcal Diff} &=& -\int dt' \, C(t-t')\frac{\partial}{\partial \sigma_i}e_{ia}(\bm{\sigma}) \nonumber \\
             ~&~& \int\left[{\mathscr D}\bm{\eta} \right]\rho(\bm{\eta})\frac{\delta}{\delta \eta_a(t')} {\bm\delta}\left(\bm{\sigma}-\bm{s}[\bm{\eta}]\right) 
 \label{Diffusion4}
\end{eqnarray}
which may be, further, simplified, by using the identities pertaining to  the Stratonovich prescription\cite{moreno2015langevin}~:

\begin{equation}
\label{Relation1}
\begin{array}{lcl}
\displaystyle
\frac{\delta}{\delta \eta_a(t')}{\bm\delta}\left(\bm{\sigma}-\bm{s}[\bm{\eta}(t)]\right)&=&\displaystyle\frac{\delta s_l(t)}{\delta \eta_a(t')}\frac{\delta}{\delta s_l(t)} {\bm\delta}\left(\bm{\sigma}-\bm{s}[\bm{\eta}(t)]\right) 
 \\
                                                                         &=&\displaystyle -\frac{\partial}{\partial \sigma_l}{\bm\delta}\left(\bm{\sigma}-\bm{s}[\bm{\eta}(t)]\right)\frac{\delta s_l(t)}{\delta \eta_a(t')} 

\end{array}
\end{equation}
Relation~(\ref{Relation1}) is then applied to equation~(\ref{Diffusion4}), and one finds~:
\begin{eqnarray}
{\mathcal Diff}&=&\int dt'\, C(t-t')\frac{\partial}{\partial \sigma_i}e_{ia}(\bm{\sigma})\frac{\partial}{\partial \sigma_l} \nonumber \\
            ~&~&\int\left[{\mathscr D}\bm{\eta} \right]\rho(\bm{\eta}) \bm{\delta}\left(\bm{\sigma}-\bm{s}[\bm{\eta}]\right)\frac{\delta s_l(t)}{\delta \eta_a(t')}  
\label{Diffusion5}
\end{eqnarray}
It is, now, necessary, to find the  expression for $\delta s_l(t) / \delta \eta_a(t')$. This may be accomplished by showing that it satisfies a differential equation, whose solution can be expressed in terms of useful quantities. 

This may be done in two steps. First, the time derivative of this term is taken~: 
\begin{eqnarray}
 \frac{\partial}{\partial t} \frac{\delta s_l(t)}{\delta\eta_a(t')} = \frac{\delta \dot{s}_l(t)}{\delta\eta_a(t')}
 \label{Derivation4}
\end{eqnarray}
Then, substituting ${\dot{s}_l(t)}$ by the RHS of the  Langevin equation one has~:
\begin{eqnarray}
 \frac{\partial}{\partial t} \frac{\delta s_l(t)}{\delta\eta_a(t')} &=&\mathfrak{A}_{lp}(t) \frac{\delta s_p(t)}{\delta \eta_a(t')}+ e_{la} (\bm{s}(t))\, \delta (t-t') \nonumber \\ 
\label{Derivation5}
\end{eqnarray}
with $\bm{\mathfrak{A}}(t)$ a matrix whose components are given by:
\begin{eqnarray}
\mathfrak{A}_{lp}(t)= \frac{\delta A_l(\bm{s}(t))}{\delta s_p(t)} + \frac{\delta e_{lb}(\bm{s}(t))}{\delta s_p(t)}\eta_b(t)  
\label{Form_A}
\end{eqnarray}
The integration of eq.(\ref{Derivation5}) requires some care because of the causal property of the Langevin equation, and, due  to the non-commuting property of the matrices that appear therein, a time-ordering operator ${\rm T}$ is necessary. Taking these facts  into account, one finds the following expression for $\delta s_l(t)/\delta\eta_a(t)$:
\begin{equation}
\label{time_order}
\frac{\delta s_l(t)}{\delta\eta_a(t')}=\varTheta(t-t')e_{ma}(\bm{s}(t))\left\lbrace \left[ {\rm T}\cdotp {e}^{~ \int\limits_{t'}^{t} dt'' \mathfrak{A}(t'') } \right]^{-1} \right\rbrace_{lm}
\end{equation}
with $\varTheta(t-t')$ the Heaviside step function, resulting from the integration over the Dirac delta function $\delta(t-t')$:
\begin{equation}
 \label{Theta}
 \varTheta(t-t')=
 \left\lbrace
 \begin{array}{lcl}
  &1& \mathrm{if}\, \,t'\in \left[0,t\right[  \\
  &1/2& \mathrm{if}\,\, t'=t  \\
  &0& \mathrm{if}\, \, t' \notin \left[0,t\right]  \\
 \end{array}
\right.
\end{equation}
The diffusion term takes, therefore, the following form~:
\begin{equation}
\label{General_diffusion}
\begin{array}{lcl}
{\mathcal Diff} &=& \displaystyle \frac{\partial}{\partial \sigma_i}e_{ia}(\bm{\sigma}) \frac{\partial}{\partial \sigma_l} e_{ma}(\bm{\sigma}) \int dt' C(t-t') \varTheta(t-t')    \\ 
   & &\displaystyle \int\left[ {\mathscr  D}\bm{\eta} \right] \rho(\bm{\eta}) \bm{\delta}\left(\bm{\sigma}-\bm{s}[\bm\eta]\right)  \left\lbrace \left[ {\rm T}\cdotp {e}^{~ \int\limits_{t'}^{t} dt'' \mathfrak{A}(t'') } \right]^{-1} \right\rbrace_{lm}  
\end{array} 
\end{equation}

Finally, we have the master equation for the probability density $P(\bm{\sigma},t)$, in the form of a continuity equation 
\begin{equation}
\frac{\partial P(\bm{\sigma},t)}{\partial t} = -\frac{\partial J_i(\bm{\sigma},t)}{\partial \sigma_i},
  \label{Master_equation_P}
\end{equation}
with $J_i(\bm{\sigma},t)$ the corresponding probability flow, given by~:
\begin{widetext}
\begin{eqnarray}
J_i(\bm{\sigma},t)&=&A_i(\bm{\sigma},\bm{\omega}) P(\bm{\sigma},t)\nonumber\\
&&-e_{ia}(\bm{\sigma}) \frac{\partial}{\partial \sigma_l} e_{ma}(\bm{\sigma}) \int dt' C(t-t') \varTheta(t-t') \int\left[ {\mathscr  D}\bm{\eta} \right] \rho(\bm{\eta}) \bm{\delta}\left(\bm{\sigma}-\bm{s}[\bm\eta]\right)  \left\lbrace \left[ {\rm T}\cdotp {e}^{~ \int\limits_{t'}^{t} dt'' \mathfrak{A}(t'') } \right]^{-1} \right\rbrace_{lm}
 \label{Probability_flow}
\end{eqnarray}
\end{widetext}
The flow term $\bm{J}$, in general, cannot be put in local form, i.e. it cannot be expressed in terms of a finite number of derivatives of a local function,  with the notable exception of a Markovian process. 

This expression is in close analogy with eq.(4) of reference \cite{Venkatesh:1993kb}, obtained here for an arbitrary target space vector field. Any further simplification of the probability flow $\bm{J}$ depends on the nature of the considered stochastic process.  Appendix (\ref{App2}) reviews  the Markovian process and how the  Fokker--Planck equation is thereby recovered from this general formalism.  In Appendix (\ref{App3}), a non-Markovian process is studied and different approximation schemes are considered to obtain a useful form for the master equation. 

\section{Fokker-Planck equation for Markovian magnetization dynamics}\label{App2}
In this appendix, we show  that the probability flow,  which , in the Markovian limit, can be expressed in local form, with finite number of derivatives in the spin variables, can be obtained from the general formalism constructed previously. 

The correlation function of a white noise process is  proportional  to a delta--function in time, $C(t-t')=2D\delta \left( t-t'\right)$. Then, the diffusion term simplifies enormously, because we only have to select the value of all the functionals into (\ref{Probability_flow}), for $t=t'$. As the time--ordering operator acts trivially for equal times, we find the following expression for the probability flow
\begin{eqnarray}
J_i(\bm{\sigma},t)&=&A_i(\bm{\sigma},\bm{\omega}) P(\bm{\sigma},t) \nonumber \\
                  &~&- D e_{ia}(\bm{\sigma})\frac{\partial}{\partial \sigma_l}\big[ e_{la}(\bm{\sigma})P(\bm{\sigma},t) \big]
\label{Probability_flow_White}
\end{eqnarray}
and, by the way, we recover the known form of the Fokker--Planck equation, valid for a manifold, parametrized by the spin variables~\cite{suhl2007relaxation,Mayergoyz:2009kl}, which is consistent with general coordinate invariance~\cite{zinn2007phase}:
\begin{equation}
\label{Fokker_Planck_White}
\begin{array}{lcl}
\displaystyle
\frac{\partial P(\bm{\sigma},t)}{\partial t}&=&\displaystyle
- \frac{\partial}{\partial \sigma_i}\Big( A_i(\bm{\sigma}) P(\bm{\sigma},t) \Big)  \\
                            ~&~&\displaystyle
                            +D\frac{\partial}{\partial \sigma_i}\left(e_{ia}(\bm{\sigma})\frac{\partial}{\partial \sigma_l}\Big( e_{la}(\bm{\sigma})P(\bm{\sigma},t) \Big) \right)     
\end{array}
\end{equation}
\section{Fokker-Planck equation for Ornstein-Uhlenbeck magnetization dynamics} \label{App3}
For a non-Markovian dynamics described by an Ornstein-Ulenbeck stochastic process, the situation is more involved--nonetheless a partial differential equation can be deduced, in the limit of weakly correlated noise. The following derivation is inspired by the work of  Fox \cite{fox1986functional,fox1986uniform} and generalized to more than one variables, which correspond to the three components of the spin. A first attempt to derive a Fokker-Planck equation for certain non-Markovian processes was given by San Miguel and Sancho \cite{san_miguel_fokker-planck_1980}, who used an expansion in $\tau$ ($\tau$ is the correlation time of the noise), to study the conditions for the existence of a well defined Fokker-Planck equation (i.e. containing first and second derivatives of the probability density only). The same conclusion was obtained  by Lindenberg and West \cite{lindenberg_finite_1984}, who proved, with the help of the Baker-Campbell-Hausdorff expansion formula, that a second-order equation with state- and time-dependent diffusion tensor exists for an arbitrary finite correlation time $\tau$. Using a partial re-summation technique of all the terms of the Fokker-Planck form,  Hänggi {\it et al.} \cite{Hanggi:1985fv}, showed that the weak noise dynamics of Fokker-Planck systems in more  than 2 state-space dimensions, is generally beset with chaotic behavior. The dynamics of such systems can be mapped onto a  non-Markovian, Langevin equation in one variable, driven by an additive, Ornstein-Ulenbeck stochastic process in a bistable potential. Such a re-summation technique was quickly generalized to multi--component systems \cite{ramirez1988higher}, but never subject to any numerical or experimental test, apparently. Moreover, the expressions obtained to date  were restricted to the special case of vanishing  diffusive kernel tensor $\Gamma$,  defined as
\begin{equation*}
\Gamma_{ijk}(\bm{\sigma})=\frac{\partial e_{ij}}{\partial\sigma_l}e_{lk}(\bm{\sigma})-\frac{\partial e_{ik}}{\partial\sigma_l}e_{lj}(\bm{\sigma}),
\end{equation*}
which doesn't vanish, in  the case of a multiplicative vielbein, which is  relevant for spins.

For such a stochastic process, the correlation function for the noise variable is
\begin{equation*}
C(t-t')=\frac{D}{\tau} e^{ -\frac{|t-t'|}{\tau} }
\end{equation*}
with $t'\in\left[ 0,t\right]$, which is, formally, equivalent to the expansion 
\begin{equation}
C(t-t')=2D\left(1+\tau\frac{d}{dt'}+\tau^2\frac{d^2}{dt'^2}+\dots\right)\delta(t-t')
\end{equation}
which does  exhibit the white-noise limit, $\lim_{\tau\rightarrow 0}C(t-t')=2D\delta(t-t')$, assuming this exists. 

Any practical application of the expressions obtained previously  requires dealing with  the time-ordered product, that  appears in (\ref{Probability_flow})~:
\begin{equation}
\left[ {\rm T}\cdotp {e}^{~ \int\limits_{t'}^{t} dt'' \mathfrak{A}(t'') } \right]^{-1}_{lm} \label{Form_deriv}
\end{equation}
For large deviations from the white-noise limit (i.e. the auto-correlation time $\tau$ takes ''large'' values), the commutator $\left[\mathfrak{A}(t), \mathfrak{A}(s) \right]$, has no reason to vanish. A relation between the Dyson perturbative series and the Magnus expansion, known from  other contexts\cite{Blanes:2009xw} can, also,  be formally obtained. However the validity of such approximations is hard to establish. 
 
For ``small'' values of $\tau$, on the other hand,  the correlation function $e^{ -\frac{|t-s|}{\tau} }$ becomes very sharply peaked, and $t$ and $s$ will take extremely close values only. In this case, the commutators of $\mathfrak{A}$ can be neglected for $|t-s|>\tau$, reducing the time--ordered product to an ordinary product. Performing a first order expansion in powers of the amplitude $\mathfrak{A}$, one has:
\begin{eqnarray}
\Delta_1=\left[{\rm T} \cdot e^{~\int\limits_{t'}^{t} dt'' \mathfrak{A}(t'')}\right]^{-1}_{lm} 
&\approx& \delta_{lm}-\int\limits_{t'}^{t} dt'' \left(\frac{\delta A_l(\bm{s}(t''))}{\delta s_m(t'')} \right.\nonumber\\
&~& \left.+ \eta_b(t) \frac{\delta e_{lb}(\bm{s}(t)}{\delta s_m(t)}\right)+\dots 
\label{A_dev3}
\end{eqnarray}
$\Delta_1$ is used in the diffusion term (\ref{General_diffusion}). The delta function $\delta_{lm}$ leads to the expression derived in the white-noise limit, whereas the two other terms, that are under the integral over $s'$, express the small deviation from the Markovian limit.

For the first correction term, the following integral has to be evaluated:
\begin{eqnarray}
\Delta_2= \int dt'~e^{~ -\frac{|t-t'|}{\tau} }\int\limits_{t'}^{t} dt'' \frac{\delta A_l(\bm{s}(t''))}{\delta s_m(t'')} 
\label{A_dev4}
\end{eqnarray}

The integral over $t'$  can, for the reasons explained above, be reduced to an integral over the interval $\left[0,t\right]$. Besides, if $t' \in\left[0,t\right]$, one also has $|t-t'|=t-t'$. 
Under the same assumptions  expression (\ref{A_dev4}) can be approximated by:
\begin{eqnarray}
\Delta_2= \frac{\delta A_l(\bm{s}(t))}{\delta s_m(t)}\int\limits_{0}^{t} dt' t' e^{-\frac{t'}{\tau} } \label{A_dev5}
\end{eqnarray}
Evaluating the integral on $t'$, and neglecting the transient terms, expression \ref{A_dev5} becomes:
\begin{eqnarray}
\Delta_2= \tau^2  \frac{\delta A_l(\bm{s}(t))}{\delta s_m(t)} \label{A_dev6}
\end{eqnarray}
Injecting $\Delta_2$ into eq.~(\ref{Probability_flow}), one has the following expression in the probability flow:
\begin{equation}
 D \tau\, e_{ia}\left( \bm{\sigma}\right) \frac{\partial}{\partial \sigma_l} e_{ma}\left( \bm{\sigma}\right) \frac{\delta A_l (\bm{\sigma},t)}{\delta \sigma_m} P(\bm{\sigma},t)
\end{equation}
 
Finally, we need to evaluate the contribution of  the third term of eq.(\ref{A_dev3}), that has the following expression:
\begin{equation}
\Delta_3=\frac{D}{\tau} \int dt' e^{ -\frac{|t-t'|}{\tau}} \int\limits_{s}^{t} dt'' \eta_b(t'') \frac{\delta e_{lb}(\bm{s}(t''))}{\delta s_m(t'')}
\label{D3_1}
\end{equation}
Computing the time integral over $t'$ (transient terms are neglected), and performing the functional derivative, one has:
 \begin{equation}
\Delta_3=2D \epsilon_{lmb} \int\limits_{s}^{t} dt'' \eta_b(t'') 
\label{D3_2}
\end{equation}
Then, eq.~(\ref{D3_2}) is injected in the probability flow equation (\ref{Probability_flow}). Applying the identities for the Levi-Civita tensors  $\epsilon_{map}\epsilon_{lmb}=\delta_{al}\delta_{pb}-\delta_{ab}\delta_{pl}$, one has:
\begin{eqnarray}
\Delta_4 &=& -2D e_{ia}(\bm{\sigma}) \left(\delta_{al}\sigma_b \frac{\partial }{\partial \sigma_l}-\delta_{ab} \sigma_l \frac{\partial }{\partial \sigma_l} \right) \nonumber \\
         &~&\int\limits_{s}^{t} dt''  \int \left[ {\mathscr D}\bm{\eta} \right] \rho(\bm{\eta}) {\bm\delta}\left(\bm{\sigma}-\bm{s}[\bm{\eta}(t)]\right)  \eta_b(t'') ~~
\label{D4_1}
\end{eqnarray}
One can, also, use  the following approximation\cite{fox1986functional,fox1986uniform,ramirez1988higher}:
\begin{equation}
 \int\limits_{s}^{t} dt'' F(t'') \approx \tau F(t)
\end{equation}
Then, denoting $K_{ab}(\bm{\sigma})= \left(\delta_{al}\sigma_b \frac{\partial }{\partial \sigma_l}-\delta_{ab} \sigma_l \frac{\partial }{\partial \sigma_l} \right)$, one has:
\begin{eqnarray}
\Delta_4 &=& -2D\tau e_{ia}K_{ab}(\bm{\sigma}) \nonumber \\
        ~&~& \int \left[ {\mathscr D}\bm{\eta} \right] \rho(\bm{\eta}) {\bm\delta}\left(\bm{\sigma}-\bm{s}[\bm{\eta}(t)]\right)  \eta_b(t) ~~
\label{D4_2}
\end{eqnarray}
Again, we apply rel.~(\ref{Distrib_deriv2}). Integrating by parts, one has:
\begin{eqnarray}
\Delta_4 &=& -2D^2 e_{ia}(\bm{\sigma})\, K_{ab}(\bm{\sigma}) \int dt' e^{ -\frac{|t-t'|}{\tau}}  \nonumber \\
        ~&~& \int \left[ {\mathscr D}\bm{\eta} \right] \rho(\bm{\eta}) {\bm\delta} \frac{\partial}{\partial  \eta_b(t')} \left(\bm{\sigma}-\bm{s}[\bm{\eta}(t)]\right)  \eta_b(t') ~~
\label{D4_3}
\end{eqnarray}
In eq.~(\ref{D4_3}), the term in the functional integral has the same form as the one in eq.~(\ref{Diffusion4}). Then, the same techniques are applied for its derivation. An expression for the functional derivative $\delta s_k(t)/\delta \eta_b(t')$ is required. Keeping, in the expression for the probability flow, only the terms that are of first order in $\tau$, one has:
\begin{equation}
\Delta_4 = 2D^2 \tau\, e_{ia}(\bm{\sigma})\, K_{ab}(\bm{\sigma}) \, \frac{\partial}{\partial \sigma_k} e_{kb}(\bm{\sigma}) P(\bm{\sigma},t)
\label{D4_4}
\end{equation}

Assembling all the terms, we, finally, obtain, for weak values of the auto-correlation time $\tau$ (i.e. slightly non-Markovian situations), an expression for the master equation, that displays the corrections from the Fokker--Planck form:
\begin{widetext}
\begin{eqnarray}
\frac{\partial P({\bm{\sigma}},t)}{\partial t}&=&- \frac{\partial}{\partial \sigma_i}\Big[ A_i(\bm{\sigma}) P(\bm{\sigma},t) \Big] +D\frac{\partial}{\partial \sigma_i}\left(
e_{ia}(\bm{\sigma})\frac{\partial}{\partial \sigma_l}\Big[ e_{la}(\bm{\sigma})P(\bm{\sigma},t) \Big]\right) \nonumber\\
&&+D\tau \frac{\partial}{\partial \sigma_i} \left(e_{ia}\left(\bm{\sigma} \right) \frac{\partial}{\partial \sigma_l} \Big[ e_{ma}\left( \bm{\sigma} \right) \frac{\delta A_l\left( \bm{\sigma} \right) }{\delta \sigma_m} P\left( \bm{\sigma},t\right) \Big]\right) \nonumber \\
                             &~& -D^2 \tau \frac{\partial}{\partial \sigma_i} \left[e_{ia}\left(\bm{\sigma} \right) \left( \delta_{al} \sigma_b \frac{\partial}{\partial \sigma_l} - \delta_{ab} \sigma_l \frac{\partial}{\partial \sigma_l}  \right)
                             \left\{ \frac{\partial}{\partial \sigma_k} \left(e_{kb}\left( \bm{\sigma}\right) P\left( \bm{\sigma},t\right)\right)\right\}\right]        
\label{Fokker_Planck_Colored}
\end{eqnarray}
\end{widetext}

A first, interesting, feature of eq.(\ref{Fokker_Planck_Colored}) is that the two first terms of its RHS are exactly the same as those of the Fokker--Planck equation, in the Markovian limit (eq.(\ref{Fokker_Planck_White}) of Appendix~\ref{App2}). Besides, when the auto-correlation time $\tau$ of the bath variables becomes negligible (i.e. $\tau \to 0$), eq.(\ref{Fokker_Planck_White}) is immediately recovered.  This is consistent with the definition of the noise correlation we chose (eq.\ref{White_noise} in Section~\ref{Intro}). 

In this sense, this new equation can be seen as an expansion about the Markovian limit, for small values of $\tau$, in the Kramers-Moyal framework  of the Fokker-Planck equation\cite{van1992stochastic}.

\bibliographystyle{apsrev4-1}
\bibliography{FP_LLB}

\begin{thebibliography}{68}%
\makeatletter
\providecommand \@ifxundefined [1]{%
 \@ifx{#1\undefined}
}%
\providecommand \@ifnum [1]{%
 \ifnum #1\expandafter \@firstoftwo
 \else \expandafter \@secondoftwo
 \fi
}%
\providecommand \@ifx [1]{%
 \ifx #1\expandafter \@firstoftwo
 \else \expandafter \@secondoftwo
 \fi
}%
\providecommand \natexlab [1]{#1}%
\providecommand \enquote  [1]{``#1''}%
\providecommand \bibnamefont  [1]{#1}%
\providecommand \bibfnamefont [1]{#1}%
\providecommand \citenamefont [1]{#1}%
\providecommand \href@noop [0]{\@secondoftwo}%
\providecommand \href [0]{\begingroup \@sanitize@url \@href}%
\providecommand \@href[1]{\@@startlink{#1}\@@href}%
\providecommand \@@href[1]{\endgroup#1\@@endlink}%
\providecommand \@sanitize@url [0]{\catcode `\\12\catcode `\$12\catcode
  `\&12\catcode `\#12\catcode `\^12\catcode `\_12\catcode `\%12\relax}%
\providecommand \@@startlink[1]{}%
\providecommand \@@endlink[0]{}%
\providecommand \url  [0]{\begingroup\@sanitize@url \@url }%
\providecommand \@url [1]{\endgroup\@href {#1}{\urlprefix }}%
\providecommand \urlprefix  [0]{URL }%
\providecommand \Eprint [0]{\href }%
\providecommand \doibase [0]{http://dx.doi.org/}%
\providecommand \selectlanguage [0]{\@gobble}%
\providecommand \bibinfo  [0]{\@secondoftwo}%
\providecommand \bibfield  [0]{\@secondoftwo}%
\providecommand \translation [1]{[#1]}%
\providecommand \BibitemOpen [0]{}%
\providecommand \bibitemStop [0]{}%
\providecommand \bibitemNoStop [0]{.\EOS\space}%
\providecommand \EOS [0]{\spacefactor3000\relax}%
\providecommand \BibitemShut  [1]{\csname bibitem#1\endcsname}%
\let\auto@bib@innerbib\@empty
\bibitem [{\citenamefont {Evans}\ \emph {et~al.}(2012)\citenamefont {Evans},
  \citenamefont {Chantrell}, \citenamefont {Nowak}, \citenamefont {Lyberatos},\
  and\ \citenamefont {Richter}}]{evans2012thermally}%
  \BibitemOpen
  \bibfield  {author} {\bibinfo {author} {\bibfnamefont {R.~F.~L.}\
  \bibnamefont {Evans}}, \bibinfo {author} {\bibfnamefont {R.~W.}\ \bibnamefont
  {Chantrell}}, \bibinfo {author} {\bibfnamefont {U.}~\bibnamefont {Nowak}},
  \bibinfo {author} {\bibfnamefont {A.}~\bibnamefont {Lyberatos}}, \ and\
  \bibinfo {author} {\bibfnamefont {H.-J.}\ \bibnamefont {Richter}},\
  }\href@noop {} {\bibfield  {journal} {\bibinfo  {journal} {Applied Physics
  Letters}\ }\textbf {\bibinfo {volume} {100}},\ \bibinfo {pages} {102402}
  (\bibinfo {year} {2012})}\BibitemShut {NoStop}%
\bibitem [{\citenamefont {Suhl}(2007)}]{suhl2007relaxation}%
  \BibitemOpen
  \bibfield  {author} {\bibinfo {author} {\bibfnamefont {H.}~\bibnamefont
  {Suhl}},\ }\href@noop {} {\emph {\bibinfo {title} {Relaxation Processes in
  Micromagnetics}}},\ \bibinfo {edition} {1st}\ ed.,\ \bibinfo {series}
  {International Series of Monograph on Physics}, Vol.\ \bibinfo {volume}
  {133}\ (\bibinfo  {publisher} {Oxford University Press},\ \bibinfo {address}
  {Oxford},\ \bibinfo {year} {2007})\BibitemShut {NoStop}%
\bibitem [{\citenamefont {Ostler}\ \emph {et~al.}(2012)\citenamefont {Ostler},
  \citenamefont {Barker}, \citenamefont {Evans}, \citenamefont {Chantrell},
  \citenamefont {Atxitia}, \citenamefont {Chubykalo-Fesenko}, \citenamefont
  {El~Moussaoui}, \citenamefont {Le~Guyader}, \citenamefont {Mengotti},
  \citenamefont {Heyderman} \emph {et~al.}}]{ostler2012ultrafast}%
  \BibitemOpen
  \bibfield  {author} {\bibinfo {author} {\bibfnamefont {T.~A.}\ \bibnamefont
  {Ostler}}, \bibinfo {author} {\bibfnamefont {J.}~\bibnamefont {Barker}},
  \bibinfo {author} {\bibfnamefont {R.~F.~L.}\ \bibnamefont {Evans}}, \bibinfo
  {author} {\bibfnamefont {R.~W.}\ \bibnamefont {Chantrell}}, \bibinfo {author}
  {\bibfnamefont {U.}~\bibnamefont {Atxitia}}, \bibinfo {author} {\bibfnamefont
  {O.}~\bibnamefont {Chubykalo-Fesenko}}, \bibinfo {author} {\bibfnamefont
  {S.}~\bibnamefont {El~Moussaoui}}, \bibinfo {author} {\bibfnamefont {L.~B.
  P.~J.}\ \bibnamefont {Le~Guyader}}, \bibinfo {author} {\bibfnamefont
  {E.}~\bibnamefont {Mengotti}}, \bibinfo {author} {\bibfnamefont {L.~J.}\
  \bibnamefont {Heyderman}},  \emph {et~al.},\ }\href@noop {} {\bibfield
  {journal} {\bibinfo  {journal} {Nature Communication}\ }\textbf {\bibinfo
  {volume} {3}},\ \bibinfo {pages} {666} (\bibinfo {year} {2012})}\BibitemShut
  {NoStop}%
\bibitem [{\citenamefont {Thiele}\ \emph {et~al.}(2003)\citenamefont {Thiele},
  \citenamefont {Maat},\ and\ \citenamefont {Fullerton}}]{thiele2003ferh}%
  \BibitemOpen
  \bibfield  {author} {\bibinfo {author} {\bibfnamefont {J.-U.}\ \bibnamefont
  {Thiele}}, \bibinfo {author} {\bibfnamefont {S.}~\bibnamefont {Maat}}, \ and\
  \bibinfo {author} {\bibfnamefont {E.~E.}\ \bibnamefont {Fullerton}},\
  }\href@noop {} {\bibfield  {journal} {\bibinfo  {journal} {Applied Physics
  Letters}\ }\textbf {\bibinfo {volume} {82}},\ \bibinfo {pages} {2859}
  (\bibinfo {year} {2003})}\BibitemShut {NoStop}%
\bibitem [{\citenamefont {N{\'e}el}(1953)}]{neel1953thermoremanent}%
  \BibitemOpen
  \bibfield  {author} {\bibinfo {author} {\bibfnamefont {L.}~\bibnamefont
  {N{\'e}el}},\ }\href@noop {} {\bibfield  {journal} {\bibinfo  {journal}
  {Reviews of Modern Physics}\ }\textbf {\bibinfo {volume} {25}},\ \bibinfo
  {pages} {293} (\bibinfo {year} {1953})}\BibitemShut {NoStop}%
\bibitem [{\citenamefont {Brown~Jr}(1979)}]{brown1979thermal}%
  \BibitemOpen
  \bibfield  {author} {\bibinfo {author} {\bibfnamefont {W.}~\bibnamefont
  {Brown~Jr}},\ }\href@noop {} {\bibfield  {journal} {\bibinfo  {journal} {IEEE
  Transactions on Magnetics}\ }\textbf {\bibinfo {volume} {15}},\ \bibinfo
  {pages} {1196} (\bibinfo {year} {1979})}\BibitemShut {NoStop}%
\bibitem [{\citenamefont {Coffey}\ and\ \citenamefont
  {Kalmykov}(2012)}]{coffey2012thermal}%
  \BibitemOpen
  \bibfield  {author} {\bibinfo {author} {\bibfnamefont {W.~T.}\ \bibnamefont
  {Coffey}}\ and\ \bibinfo {author} {\bibfnamefont {Y.~P.}\ \bibnamefont
  {Kalmykov}},\ }\href@noop {} {\bibfield  {journal} {\bibinfo  {journal}
  {Journal of Applied Physics}\ }\textbf {\bibinfo {volume} {112}},\ \bibinfo
  {pages} {121301} (\bibinfo {year} {2012})}\BibitemShut {NoStop}%
\bibitem [{\citenamefont {Gardiner}(1985)}]{gardiner1985stochastic}%
  \BibitemOpen
  \bibfield  {author} {\bibinfo {author} {\bibfnamefont {C.~W.}\ \bibnamefont
  {Gardiner}},\ }\href@noop {} {\emph {\bibinfo {title} {Handbook of Stochastic
  Methods}}},\ \bibinfo {edition} {2nd}\ ed.\ (\bibinfo  {publisher}
  {Springer-Verlag},\ \bibinfo {address} {Berlin--Heidelberg--New
  York--Tokyo},\ \bibinfo {year} {1985})\BibitemShut {NoStop}%
\bibitem [{\citenamefont {Van~Kampen}(2007)}]{van1992stochastic}%
  \BibitemOpen
  \bibfield  {author} {\bibinfo {author} {\bibfnamefont {N.~G.}\ \bibnamefont
  {Van~Kampen}},\ }\href@noop {} {\emph {\bibinfo {title} {Stochastic Processes
  in Physics and Chemistry}}},\ \bibinfo {edition} {3rd}\ ed.,\ Vol.~\bibinfo
  {volume} {1}\ (\bibinfo  {publisher} {Elsevier},\ \bibinfo {address}
  {Amsterdam},\ \bibinfo {year} {2007})\BibitemShut {NoStop}%
\bibitem [{\citenamefont {Simon}\ \emph {et~al.}(2014)\citenamefont {Simon},
  \citenamefont {Palot{\'a}s}, \citenamefont {Ujfalussy}, \citenamefont
  {De{\'a}k}, \citenamefont {Stocks},\ and\ \citenamefont
  {Szunyogh}}]{simon2014spin}%
  \BibitemOpen
  \bibfield  {author} {\bibinfo {author} {\bibfnamefont {E.}~\bibnamefont
  {Simon}}, \bibinfo {author} {\bibfnamefont {K.}~\bibnamefont {Palot{\'a}s}},
  \bibinfo {author} {\bibfnamefont {B.}~\bibnamefont {Ujfalussy}}, \bibinfo
  {author} {\bibfnamefont {A.}~\bibnamefont {De{\'a}k}}, \bibinfo {author}
  {\bibfnamefont {G.~M.}\ \bibnamefont {Stocks}}, \ and\ \bibinfo {author}
  {\bibfnamefont {L.}~\bibnamefont {Szunyogh}},\ }\href@noop {} {\bibfield
  {journal} {\bibinfo  {journal} {Journal of Physics: Condensed Matter}\
  }\textbf {\bibinfo {volume} {26}},\ \bibinfo {pages} {186001} (\bibinfo
  {year} {2014})}\BibitemShut {NoStop}%
\bibitem [{\citenamefont {Tranchida}\ \emph {et~al.}(2016)\citenamefont
  {Tranchida}, \citenamefont {Thibaudeau},\ and\ \citenamefont
  {Nicolis}}]{tranchida_closing_2016}%
  \BibitemOpen
  \bibfield  {author} {\bibinfo {author} {\bibfnamefont {J.}~\bibnamefont
  {Tranchida}}, \bibinfo {author} {\bibfnamefont {P.}~\bibnamefont
  {Thibaudeau}}, \ and\ \bibinfo {author} {\bibfnamefont {S.}~\bibnamefont
  {Nicolis}},\ }\href {\doibase 10.1016/j.physb.2015.10.012} {\bibfield
  {journal} {\bibinfo  {journal} {Physica B: Condensed Matter}\ }\textbf
  {\bibinfo {volume} {486}},\ \bibinfo {pages} {57} (\bibinfo {year}
  {2016})}\BibitemShut {NoStop}%
\bibitem [{\citenamefont {Risken}(1984)}]{risken1984fokker}%
  \BibitemOpen
  \bibfield  {author} {\bibinfo {author} {\bibfnamefont {H.}~\bibnamefont
  {Risken}},\ }\href@noop {} {\emph {\bibinfo {title} {The Fokker-Planck
  Equation}}},\ \bibinfo {edition} {2nd}\ ed.,\ \bibinfo {series} {Springer
  Series in Synergetics}, Vol.~\bibinfo {volume} {18}\ (\bibinfo  {publisher}
  {Springer-Verlag},\ \bibinfo {address} {Berlin--Heidelberg--New
  York--Tokyo},\ \bibinfo {year} {1984})\BibitemShut {NoStop}%
\bibitem [{\citenamefont {Mayergoyz}\ \emph {et~al.}(2009)\citenamefont
  {Mayergoyz}, \citenamefont {Bertotti},\ and\ \citenamefont
  {Serpico}}]{Mayergoyz:2009kl}%
  \BibitemOpen
  \bibfield  {author} {\bibinfo {author} {\bibfnamefont {I.~D.}\ \bibnamefont
  {Mayergoyz}}, \bibinfo {author} {\bibfnamefont {G.}~\bibnamefont {Bertotti}},
  \ and\ \bibinfo {author} {\bibfnamefont {C.}~\bibnamefont {Serpico}},\
  }\href@noop {} {\emph {\bibinfo {title} {Nonlinear Magnetization Dynamics in
  Nanosystems}}},\ \bibinfo {edition} {1st}\ ed.,\ Elsevier Series in
  Electromagnetism\ (\bibinfo  {publisher} {Elsevier},\ \bibinfo {address}
  {Amsterdam},\ \bibinfo {year} {2009})\BibitemShut {NoStop}%
\bibitem [{\citenamefont {Evans}\ \emph {et~al.}(2014)\citenamefont {Evans},
  \citenamefont {Fan}, \citenamefont {Chureemart}, \citenamefont {Ostler},
  \citenamefont {Ellis},\ and\ \citenamefont {Chantrell}}]{evans2014atomistic}%
  \BibitemOpen
  \bibfield  {author} {\bibinfo {author} {\bibfnamefont {R.~F.~L.}\
  \bibnamefont {Evans}}, \bibinfo {author} {\bibfnamefont {W.~J.}\ \bibnamefont
  {Fan}}, \bibinfo {author} {\bibfnamefont {P.}~\bibnamefont {Chureemart}},
  \bibinfo {author} {\bibfnamefont {T.~A.}\ \bibnamefont {Ostler}}, \bibinfo
  {author} {\bibfnamefont {M.~O.~A.}\ \bibnamefont {Ellis}}, \ and\ \bibinfo
  {author} {\bibfnamefont {R.~W.}\ \bibnamefont {Chantrell}},\ }\href@noop {}
  {\bibfield  {journal} {\bibinfo  {journal} {Journal of Physics: Condensed
  Matter}\ }\textbf {\bibinfo {volume} {26}},\ \bibinfo {pages} {103202}
  (\bibinfo {year} {2014})}\BibitemShut {NoStop}%
\bibitem [{\citenamefont {Beaujouan}\ \emph
  {et~al.}(2012{\natexlab{a}})\citenamefont {Beaujouan}, \citenamefont
  {Thibaudeau},\ and\ \citenamefont {Barreteau}}]{beaujouan2012anisotropic}%
  \BibitemOpen
  \bibfield  {author} {\bibinfo {author} {\bibfnamefont {D.}~\bibnamefont
  {Beaujouan}}, \bibinfo {author} {\bibfnamefont {P.}~\bibnamefont
  {Thibaudeau}}, \ and\ \bibinfo {author} {\bibfnamefont {C.}~\bibnamefont
  {Barreteau}},\ }\href@noop {} {\bibfield  {journal} {\bibinfo  {journal}
  {Physical Review B}\ }\textbf {\bibinfo {volume} {86}},\ \bibinfo {pages}
  {174409} (\bibinfo {year} {2012}{\natexlab{a}})}\BibitemShut {NoStop}%
\bibitem [{\citenamefont {Garanin}\ \emph {et~al.}(1990)\citenamefont
  {Garanin}, \citenamefont {Ishchenko},\ and\ \citenamefont
  {Panina}}]{garanin1990dynamics}%
  \BibitemOpen
  \bibfield  {author} {\bibinfo {author} {\bibfnamefont {D.~A.}\ \bibnamefont
  {Garanin}}, \bibinfo {author} {\bibfnamefont {V.~V.}\ \bibnamefont
  {Ishchenko}}, \ and\ \bibinfo {author} {\bibfnamefont {L.~V.}\ \bibnamefont
  {Panina}},\ }\href@noop {} {\bibfield  {journal} {\bibinfo  {journal}
  {Theoretical and Mathematical Physics}\ }\textbf {\bibinfo {volume} {82}},\
  \bibinfo {pages} {169} (\bibinfo {year} {1990})}\BibitemShut {NoStop}%
\bibitem [{\citenamefont {Garanin}(1997)}]{garanin1997fokker}%
  \BibitemOpen
  \bibfield  {author} {\bibinfo {author} {\bibfnamefont {D.~A.}\ \bibnamefont
  {Garanin}},\ }\href@noop {} {\bibfield  {journal} {\bibinfo  {journal}
  {Physical Review B}\ }\textbf {\bibinfo {volume} {55}},\ \bibinfo {pages}
  {3050} (\bibinfo {year} {1997})}\BibitemShut {NoStop}%
\bibitem [{\citenamefont {Kazantseva}\ \emph {et~al.}(2008)\citenamefont
  {Kazantseva}, \citenamefont {Hinzke}, \citenamefont {Nowak}, \citenamefont
  {Chantrell}, \citenamefont {Atxitia},\ and\ \citenamefont
  {Chubykalo-Fesenko}}]{kazantseva2008towards}%
  \BibitemOpen
  \bibfield  {author} {\bibinfo {author} {\bibfnamefont {N.}~\bibnamefont
  {Kazantseva}}, \bibinfo {author} {\bibfnamefont {D.}~\bibnamefont {Hinzke}},
  \bibinfo {author} {\bibfnamefont {U.}~\bibnamefont {Nowak}}, \bibinfo
  {author} {\bibfnamefont {R.~W.}\ \bibnamefont {Chantrell}}, \bibinfo {author}
  {\bibfnamefont {U.}~\bibnamefont {Atxitia}}, \ and\ \bibinfo {author}
  {\bibfnamefont {O.}~\bibnamefont {Chubykalo-Fesenko}},\ }\href@noop {}
  {\bibfield  {journal} {\bibinfo  {journal} {Physical Review B}\ }\textbf
  {\bibinfo {volume} {77}},\ \bibinfo {pages} {184428} (\bibinfo {year}
  {2008})}\BibitemShut {NoStop}%
\bibitem [{\citenamefont {Bouchaud}\ and\ \citenamefont
  {Z{\'e}rah}(1989)}]{bouchaud1989spontaneous}%
  \BibitemOpen
  \bibfield  {author} {\bibinfo {author} {\bibfnamefont {J.~P.}\ \bibnamefont
  {Bouchaud}}\ and\ \bibinfo {author} {\bibfnamefont {P.~G.}\ \bibnamefont
  {Z{\'e}rah}},\ }\href@noop {} {\bibfield  {journal} {\bibinfo  {journal}
  {Physical Review Letters}\ }\textbf {\bibinfo {volume} {63}},\ \bibinfo
  {pages} {1000} (\bibinfo {year} {1989})}\BibitemShut {NoStop}%
\bibitem [{\citenamefont {Thibaudeau}\ and\ \citenamefont
  {Tranchida}(2015)}]{thibaudeau2015frequency}%
  \BibitemOpen
  \bibfield  {author} {\bibinfo {author} {\bibfnamefont {P.}~\bibnamefont
  {Thibaudeau}}\ and\ \bibinfo {author} {\bibfnamefont {J.}~\bibnamefont
  {Tranchida}},\ }\href@noop {} {\bibfield  {journal} {\bibinfo  {journal}
  {Journal of Applied Physics}\ }\textbf {\bibinfo {volume} {118}},\ \bibinfo
  {pages} {053901} (\bibinfo {year} {2015})}\BibitemShut {NoStop}%
\bibitem [{\citenamefont {Zinn-Justin}(2002)}]{justin1989quantum}%
  \BibitemOpen
  \bibfield  {author} {\bibinfo {author} {\bibfnamefont {J.}~\bibnamefont
  {Zinn-Justin}},\ }\href {\doibase 10.1093/acprof:050/9780198509233.001.0001}
  {\emph {\bibinfo {title} {Quantum Field Theory and Critical Phenomena}}},\
  \bibinfo {edition} {4th}\ ed.,\ \bibinfo {series} {International Series of
  Monograph on Physics}, Vol.\ \bibinfo {volume} {113}\ (\bibinfo  {publisher}
  {Oxford University Press},\ \bibinfo {address} {Oxford},\ \bibinfo {year}
  {2002})\BibitemShut {NoStop}%
\bibitem [{\citenamefont {Zinn-Justin}(2007)}]{zinn2007phase}%
  \BibitemOpen
  \bibfield  {author} {\bibinfo {author} {\bibfnamefont {J.}~\bibnamefont
  {Zinn-Justin}},\ }\href@noop {} {\emph {\bibinfo {title} {Phase Transitions
  and Renormalization Group}}},\ \bibinfo {edition} {1st}\ ed.,\ Oxford
  Graduate Texts\ (\bibinfo  {publisher} {Oxford University Press},\ \bibinfo
  {address} {Oxford},\ \bibinfo {year} {2007})\BibitemShut {NoStop}%
\bibitem [{\citenamefont {Kleinert}(2009)}]{kleinert2009path}%
  \BibitemOpen
  \bibfield  {author} {\bibinfo {author} {\bibfnamefont {H.}~\bibnamefont
  {Kleinert}},\ }\href@noop {} {\emph {\bibinfo {title} {Path {Integrals} in
  {Quantum} {Mechanics}, {Statistics}, {Polymer} {Physics}, and {Financial}
  {Markets}}}},\ \bibinfo {edition} {3rd}\ ed.\ (\bibinfo  {publisher} {World
  Scientific},\ \bibinfo {year} {2009})\BibitemShut {NoStop}%
\bibitem [{\citenamefont {Aron}\ \emph {et~al.}(2014)\citenamefont {Aron},
  \citenamefont {Barci}, \citenamefont {Cugliandolo}, \citenamefont {Arenas},\
  and\ \citenamefont {Lozano}}]{aron2014magnetization}%
  \BibitemOpen
  \bibfield  {author} {\bibinfo {author} {\bibfnamefont {C.}~\bibnamefont
  {Aron}}, \bibinfo {author} {\bibfnamefont {D.~G.}\ \bibnamefont {Barci}},
  \bibinfo {author} {\bibfnamefont {L.~F.}\ \bibnamefont {Cugliandolo}},
  \bibinfo {author} {\bibfnamefont {Z.~G.}\ \bibnamefont {Arenas}}, \ and\
  \bibinfo {author} {\bibfnamefont {G.~S.}\ \bibnamefont {Lozano}},\
  }\href@noop {} {\bibfield  {journal} {\bibinfo  {journal} {Journal of
  Statistical Mechanics: Theory and Experiment}\ }\textbf {\bibinfo {volume}
  {2014}},\ \bibinfo {pages} {P09008} (\bibinfo {year} {2014})}\BibitemShut
  {NoStop}%
\bibitem [{\citenamefont {Moreno}\ \emph {et~al.}(2015)\citenamefont {Moreno},
  \citenamefont {Arenas},\ and\ \citenamefont {Barci}}]{moreno2015langevin}%
  \BibitemOpen
  \bibfield  {author} {\bibinfo {author} {\bibfnamefont {M.~V.}\ \bibnamefont
  {Moreno}}, \bibinfo {author} {\bibfnamefont {Z.~G.}\ \bibnamefont {Arenas}},
  \ and\ \bibinfo {author} {\bibfnamefont {D.~G.}\ \bibnamefont {Barci}},\
  }\href@noop {} {\bibfield  {journal} {\bibinfo  {journal} {Physical Review
  E}\ }\textbf {\bibinfo {volume} {91}},\ \bibinfo {pages} {042103} (\bibinfo
  {year} {2015})}\BibitemShut {NoStop}%
\bibitem [{\citenamefont {Beaurepaire}\ \emph {et~al.}(1996)\citenamefont
  {Beaurepaire}, \citenamefont {Merle}, \citenamefont {Daunois},\ and\
  \citenamefont {Bigot}}]{beaurepaire1996ultrafast}%
  \BibitemOpen
  \bibfield  {author} {\bibinfo {author} {\bibfnamefont {E.}~\bibnamefont
  {Beaurepaire}}, \bibinfo {author} {\bibfnamefont {J.-C.}\ \bibnamefont
  {Merle}}, \bibinfo {author} {\bibfnamefont {A.}~\bibnamefont {Daunois}}, \
  and\ \bibinfo {author} {\bibfnamefont {J.-Y.}\ \bibnamefont {Bigot}},\
  }\href@noop {} {\bibfield  {journal} {\bibinfo  {journal} {Physical Review
  Letters}\ }\textbf {\bibinfo {volume} {76}},\ \bibinfo {pages} {4250}
  (\bibinfo {year} {1996})}\BibitemShut {NoStop}%
\bibitem [{\citenamefont {Ounadjela}\ and\ \citenamefont
  {Hillebrands}(2003)}]{ounadjela2003spin}%
  \BibitemOpen
  \bibfield  {author} {\bibinfo {author} {\bibfnamefont {K.}~\bibnamefont
  {Ounadjela}}\ and\ \bibinfo {author} {\bibfnamefont {B.}~\bibnamefont
  {Hillebrands}},\ }\href@noop {} {\emph {\bibinfo {title} {Spin Dynamics in
  Confined Magnetic Structures II}}},\ \bibinfo {series} {Topics in Applied
  Physics}, Vol.~\bibinfo {volume} {87}\ (\bibinfo  {publisher} {Springer},\
  \bibinfo {address} {Berlin--Heidelberg--New York--Tokyo},\ \bibinfo {year}
  {2003})\BibitemShut {NoStop}%
\bibitem [{\citenamefont {Fox}\ \emph {et~al.}(1988)\citenamefont {Fox},
  \citenamefont {Gatland}, \citenamefont {Roy},\ and\ \citenamefont
  {Vemuri}}]{fox1988fast}%
  \BibitemOpen
  \bibfield  {author} {\bibinfo {author} {\bibfnamefont {R.~F.}\ \bibnamefont
  {Fox}}, \bibinfo {author} {\bibfnamefont {I.~R.}\ \bibnamefont {Gatland}},
  \bibinfo {author} {\bibfnamefont {R.}~\bibnamefont {Roy}}, \ and\ \bibinfo
  {author} {\bibfnamefont {G.}~\bibnamefont {Vemuri}},\ }\href@noop {}
  {\bibfield  {journal} {\bibinfo  {journal} {Physical Review A}\ }\textbf
  {\bibinfo {volume} {38}},\ \bibinfo {pages} {5938} (\bibinfo {year}
  {1988})}\BibitemShut {NoStop}%
\bibitem [{\citenamefont {Hanggi}\ and\ \citenamefont
  {Jung}(1995)}]{hanggi1995colored}%
  \BibitemOpen
  \bibfield  {author} {\bibinfo {author} {\bibfnamefont {P.}~\bibnamefont
  {Hanggi}}\ and\ \bibinfo {author} {\bibfnamefont {P.}~\bibnamefont {Jung}},\
  }\href@noop {} {\bibfield  {journal} {\bibinfo  {journal} {Advances in
  Chemical Physics}\ }\textbf {\bibinfo {volume} {89}},\ \bibinfo {pages} {239}
  (\bibinfo {year} {1995})}\BibitemShut {NoStop}%
\bibitem [{\citenamefont {Bose}\ and\ \citenamefont
  {Trimper}(2010)}]{bose2010correlation}%
  \BibitemOpen
  \bibfield  {author} {\bibinfo {author} {\bibfnamefont {T.}~\bibnamefont
  {Bose}}\ and\ \bibinfo {author} {\bibfnamefont {S.}~\bibnamefont {Trimper}},\
  }\href@noop {} {\bibfield  {journal} {\bibinfo  {journal} {Physical Review
  B}\ }\textbf {\bibinfo {volume} {81}},\ \bibinfo {pages} {104413} (\bibinfo
  {year} {2010})}\BibitemShut {NoStop}%
\bibitem [{\citenamefont {Miyazaki}\ and\ \citenamefont
  {Seki}(1998)}]{miyazaki1998brownian}%
  \BibitemOpen
  \bibfield  {author} {\bibinfo {author} {\bibfnamefont {K.}~\bibnamefont
  {Miyazaki}}\ and\ \bibinfo {author} {\bibfnamefont {K.}~\bibnamefont
  {Seki}},\ }\href@noop {} {\bibfield  {journal} {\bibinfo  {journal} {The
  Journal of Chemical Physics}\ }\textbf {\bibinfo {volume} {108}},\ \bibinfo
  {pages} {7052} (\bibinfo {year} {1998})}\BibitemShut {NoStop}%
\bibitem [{\citenamefont {Atxitia}\ \emph {et~al.}(2009)\citenamefont
  {Atxitia}, \citenamefont {Chubykalo-Fesenko}, \citenamefont {Chantrell},
  \citenamefont {Nowak},\ and\ \citenamefont {Rebei}}]{atxitia2009ultrafast}%
  \BibitemOpen
  \bibfield  {author} {\bibinfo {author} {\bibfnamefont {U.}~\bibnamefont
  {Atxitia}}, \bibinfo {author} {\bibfnamefont {O.}~\bibnamefont
  {Chubykalo-Fesenko}}, \bibinfo {author} {\bibfnamefont {R.~W.}\ \bibnamefont
  {Chantrell}}, \bibinfo {author} {\bibfnamefont {U.}~\bibnamefont {Nowak}}, \
  and\ \bibinfo {author} {\bibfnamefont {A.}~\bibnamefont {Rebei}},\
  }\href@noop {} {\bibfield  {journal} {\bibinfo  {journal} {Physical Review
  Letters}\ }\textbf {\bibinfo {volume} {102}},\ \bibinfo {pages} {057203}
  (\bibinfo {year} {2009})}\BibitemShut {NoStop}%
\bibitem [{\citenamefont {Tranchida}\ \emph {et~al.}(2015)\citenamefont
  {Tranchida}, \citenamefont {Thibaudeau},\ and\ \citenamefont
  {Nicolis}}]{tranchida2015colored}%
  \BibitemOpen
  \bibfield  {author} {\bibinfo {author} {\bibfnamefont {J.}~\bibnamefont
  {Tranchida}}, \bibinfo {author} {\bibfnamefont {P.}~\bibnamefont
  {Thibaudeau}}, \ and\ \bibinfo {author} {\bibfnamefont {S.}~\bibnamefont
  {Nicolis}},\ }\href@noop {} {\bibfield  {journal} {\bibinfo  {journal} {arXiv
  preprint arXiv:1511.02008}\ } (\bibinfo {year} {2015})}\BibitemShut {NoStop}%
\bibitem [{\citenamefont {Rom{\'a}}\ \emph {et~al.}(2014)\citenamefont
  {Rom{\'a}}, \citenamefont {Cugliandolo},\ and\ \citenamefont
  {Lozano}}]{roma2014numerical}%
  \BibitemOpen
  \bibfield  {author} {\bibinfo {author} {\bibfnamefont {F.}~\bibnamefont
  {Rom{\'a}}}, \bibinfo {author} {\bibfnamefont {L.~F.}\ \bibnamefont
  {Cugliandolo}}, \ and\ \bibinfo {author} {\bibfnamefont {G.~S.}\ \bibnamefont
  {Lozano}},\ }\href@noop {} {\bibfield  {journal} {\bibinfo  {journal}
  {Physical Review E}\ }\textbf {\bibinfo {volume} {90}},\ \bibinfo {pages}
  {023203} (\bibinfo {year} {2014})}\BibitemShut {NoStop}%
\bibitem [{\citenamefont {d'Aquino}\ \emph {et~al.}(2005)\citenamefont
  {d'Aquino}, \citenamefont {Serpico},\ and\ \citenamefont
  {Miano}}]{d2005geometrical}%
  \BibitemOpen
  \bibfield  {author} {\bibinfo {author} {\bibfnamefont {M.}~\bibnamefont
  {d'Aquino}}, \bibinfo {author} {\bibfnamefont {C.}~\bibnamefont {Serpico}}, \
  and\ \bibinfo {author} {\bibfnamefont {G.}~\bibnamefont {Miano}},\
  }\href@noop {} {\bibfield  {journal} {\bibinfo  {journal} {Journal of
  Computational Physics}\ }\textbf {\bibinfo {volume} {209}},\ \bibinfo {pages}
  {730} (\bibinfo {year} {2005})}\BibitemShut {NoStop}%
\bibitem [{\citenamefont {Skubic}\ \emph {et~al.}(2008)\citenamefont {Skubic},
  \citenamefont {Hellsvik}, \citenamefont {Nordstr{\"o}m},\ and\ \citenamefont
  {Eriksson}}]{skubic2008method}%
  \BibitemOpen
  \bibfield  {author} {\bibinfo {author} {\bibfnamefont {B.}~\bibnamefont
  {Skubic}}, \bibinfo {author} {\bibfnamefont {J.}~\bibnamefont {Hellsvik}},
  \bibinfo {author} {\bibfnamefont {L.}~\bibnamefont {Nordstr{\"o}m}}, \ and\
  \bibinfo {author} {\bibfnamefont {O.}~\bibnamefont {Eriksson}},\ }\href@noop
  {} {\bibfield  {journal} {\bibinfo  {journal} {Journal of Physics: Condensed
  Matter}\ }\textbf {\bibinfo {volume} {20}},\ \bibinfo {pages} {315203}
  (\bibinfo {year} {2008})}\BibitemShut {NoStop}%
\bibitem [{\citenamefont {Anderson}\ and\ \citenamefont
  {Weiss}(1953)}]{anderson1953exchange}%
  \BibitemOpen
  \bibfield  {author} {\bibinfo {author} {\bibfnamefont {P.-W.}\ \bibnamefont
  {Anderson}}\ and\ \bibinfo {author} {\bibfnamefont {P.~R.}\ \bibnamefont
  {Weiss}},\ }\href@noop {} {\bibfield  {journal} {\bibinfo  {journal} {Reviews
  of Modern Physics}\ }\textbf {\bibinfo {volume} {25}},\ \bibinfo {pages}
  {269} (\bibinfo {year} {1953})}\BibitemShut {NoStop}%
\bibitem [{\citenamefont {Reimers}\ \emph {et~al.}(1991)\citenamefont
  {Reimers}, \citenamefont {Berlinsky},\ and\ \citenamefont
  {Shi}}]{reimers1991mean}%
  \BibitemOpen
  \bibfield  {author} {\bibinfo {author} {\bibfnamefont {J.~N.}\ \bibnamefont
  {Reimers}}, \bibinfo {author} {\bibfnamefont {A.~J.}\ \bibnamefont
  {Berlinsky}}, \ and\ \bibinfo {author} {\bibfnamefont {A.-C.}\ \bibnamefont
  {Shi}},\ }\href@noop {} {\bibfield  {journal} {\bibinfo  {journal} {Physical
  Review B}\ }\textbf {\bibinfo {volume} {43}},\ \bibinfo {pages} {865}
  (\bibinfo {year} {1991})}\BibitemShut {NoStop}%
\bibitem [{\citenamefont {Frisch}(1995)}]{frisch1995turbulence}%
  \BibitemOpen
  \bibfield  {author} {\bibinfo {author} {\bibfnamefont {U.}~\bibnamefont
  {Frisch}},\ }\href@noop {} {\emph {\bibinfo {title} {Turbulence: the legacy
  of AN Kolmogorov}}}\ (\bibinfo  {publisher} {Cambridge university press},\
  \bibinfo {year} {1995})\BibitemShut {NoStop}%
\bibitem [{\citenamefont {Mellor}\ and\ \citenamefont
  {Yamada}(1974)}]{mellor1974hierarchy}%
  \BibitemOpen
  \bibfield  {author} {\bibinfo {author} {\bibfnamefont {G.~L.}\ \bibnamefont
  {Mellor}}\ and\ \bibinfo {author} {\bibfnamefont {T.}~\bibnamefont
  {Yamada}},\ }\href@noop {} {\bibfield  {journal} {\bibinfo  {journal}
  {Journal of the Atmospheric Sciences}\ }\textbf {\bibinfo {volume} {31}},\
  \bibinfo {pages} {1791} (\bibinfo {year} {1974})}\BibitemShut {NoStop}%
\bibitem [{\citenamefont {Nicolis}(2012)}]{nicolis2012dynamics}%
  \BibitemOpen
  \bibfield  {author} {\bibinfo {author} {\bibfnamefont {J.~S.}\ \bibnamefont
  {Nicolis}},\ }\href@noop {} {\emph {\bibinfo {title} {Dynamics of
  hierarchical systems: an evolutionary approach}}},\ Vol.~\bibinfo {volume}
  {25}\ (\bibinfo  {publisher} {Springer Science \& Business Media},\ \bibinfo
  {year} {2012})\BibitemShut {NoStop}%
\bibitem [{\citenamefont {Antropov}\ \emph {et~al.}(1995)\citenamefont
  {Antropov}, \citenamefont {Katsnelson}, \citenamefont {van Schilfgaarde},\
  and\ \citenamefont {Harmon}}]{antropov1995ab}%
  \BibitemOpen
  \bibfield  {author} {\bibinfo {author} {\bibfnamefont {V.~P.}\ \bibnamefont
  {Antropov}}, \bibinfo {author} {\bibfnamefont {M.~I.}\ \bibnamefont
  {Katsnelson}}, \bibinfo {author} {\bibfnamefont {M.}~\bibnamefont {van
  Schilfgaarde}}, \ and\ \bibinfo {author} {\bibfnamefont {B.~N.}\ \bibnamefont
  {Harmon}},\ }\href@noop {} {\bibfield  {journal} {\bibinfo  {journal}
  {Physical review letters}\ }\textbf {\bibinfo {volume} {75}},\ \bibinfo
  {pages} {729} (\bibinfo {year} {1995})}\BibitemShut {NoStop}%
\bibitem [{\citenamefont {Antropov}\ \emph {et~al.}(1996)\citenamefont
  {Antropov}, \citenamefont {Katsnelson}, \citenamefont {Harmon}, \citenamefont
  {van Schilfgaarde},\ and\ \citenamefont {Kusnezov}}]{antropov1996spin}%
  \BibitemOpen
  \bibfield  {author} {\bibinfo {author} {\bibfnamefont {V.~P.}\ \bibnamefont
  {Antropov}}, \bibinfo {author} {\bibfnamefont {M.~I.}\ \bibnamefont
  {Katsnelson}}, \bibinfo {author} {\bibfnamefont {B.~N.}\ \bibnamefont
  {Harmon}}, \bibinfo {author} {\bibfnamefont {M.}~\bibnamefont {van
  Schilfgaarde}}, \ and\ \bibinfo {author} {\bibfnamefont {D.}~\bibnamefont
  {Kusnezov}},\ }\href@noop {} {\bibfield  {journal} {\bibinfo  {journal}
  {Physical Review B}\ }\textbf {\bibinfo {volume} {54}},\ \bibinfo {pages}
  {1019} (\bibinfo {year} {1996})}\BibitemShut {NoStop}%
\bibitem [{\citenamefont {Nowak}\ \emph {et~al.}(2005)\citenamefont {Nowak},
  \citenamefont {Mryasov}, \citenamefont {Wieser}, \citenamefont {Guslienko},\
  and\ \citenamefont {Chantrell}}]{nowak2005spin}%
  \BibitemOpen
  \bibfield  {author} {\bibinfo {author} {\bibfnamefont {U.}~\bibnamefont
  {Nowak}}, \bibinfo {author} {\bibfnamefont {O.~N.}\ \bibnamefont {Mryasov}},
  \bibinfo {author} {\bibfnamefont {R.}~\bibnamefont {Wieser}}, \bibinfo
  {author} {\bibfnamefont {K.}~\bibnamefont {Guslienko}}, \ and\ \bibinfo
  {author} {\bibfnamefont {R.~W.}\ \bibnamefont {Chantrell}},\ }\href@noop {}
  {\bibfield  {journal} {\bibinfo  {journal} {Physical Review B}\ }\textbf
  {\bibinfo {volume} {72}},\ \bibinfo {pages} {172410} (\bibinfo {year}
  {2005})}\BibitemShut {NoStop}%
\bibitem [{\citenamefont {Krech}\ \emph {et~al.}(1998)\citenamefont {Krech},
  \citenamefont {Bunker},\ and\ \citenamefont {Landau}}]{krech1998fast}%
  \BibitemOpen
  \bibfield  {author} {\bibinfo {author} {\bibfnamefont {M.}~\bibnamefont
  {Krech}}, \bibinfo {author} {\bibfnamefont {A.}~\bibnamefont {Bunker}}, \
  and\ \bibinfo {author} {\bibfnamefont {D.~P.}\ \bibnamefont {Landau}},\
  }\href@noop {} {\bibfield  {journal} {\bibinfo  {journal} {Computer physics
  communications}\ }\textbf {\bibinfo {volume} {111}},\ \bibinfo {pages} {1}
  (\bibinfo {year} {1998})}\BibitemShut {NoStop}%
\bibitem [{\citenamefont {Omelyan}\ \emph {et~al.}(2003)\citenamefont
  {Omelyan}, \citenamefont {Mryglod},\ and\ \citenamefont
  {Folk}}]{omelyan2003symplectic}%
  \BibitemOpen
  \bibfield  {author} {\bibinfo {author} {\bibfnamefont {I.~P.}\ \bibnamefont
  {Omelyan}}, \bibinfo {author} {\bibfnamefont {I.~M.}\ \bibnamefont
  {Mryglod}}, \ and\ \bibinfo {author} {\bibfnamefont {R.}~\bibnamefont
  {Folk}},\ }\href@noop {} {\bibfield  {journal} {\bibinfo  {journal} {Computer
  Physics Communications}\ }\textbf {\bibinfo {volume} {151}},\ \bibinfo
  {pages} {272} (\bibinfo {year} {2003})}\BibitemShut {NoStop}%
\bibitem [{\citenamefont {Ma}\ \emph {et~al.}(2008)\citenamefont {Ma},
  \citenamefont {Woo},\ and\ \citenamefont {Dudarev}}]{ma2008large}%
  \BibitemOpen
  \bibfield  {author} {\bibinfo {author} {\bibfnamefont {P.-W.}\ \bibnamefont
  {Ma}}, \bibinfo {author} {\bibfnamefont {C.~H.}\ \bibnamefont {Woo}}, \ and\
  \bibinfo {author} {\bibfnamefont {S.~L.}\ \bibnamefont {Dudarev}},\
  }\href@noop {} {\bibfield  {journal} {\bibinfo  {journal} {Physical Review
  B}\ }\textbf {\bibinfo {volume} {78}},\ \bibinfo {pages} {024434} (\bibinfo
  {year} {2008})}\BibitemShut {NoStop}%
\bibitem [{\citenamefont {Beaujouan}\ \emph
  {et~al.}(2012{\natexlab{b}})\citenamefont {Beaujouan}, \citenamefont
  {Thibaudeau},\ and\ \citenamefont {Barreteau}}]{beaujouan2012thermal}%
  \BibitemOpen
  \bibfield  {author} {\bibinfo {author} {\bibfnamefont {D.}~\bibnamefont
  {Beaujouan}}, \bibinfo {author} {\bibfnamefont {P.}~\bibnamefont
  {Thibaudeau}}, \ and\ \bibinfo {author} {\bibfnamefont {C.}~\bibnamefont
  {Barreteau}},\ }\href@noop {} {\bibfield  {journal} {\bibinfo  {journal}
  {Journal of Applied Physics}\ }\textbf {\bibinfo {volume} {111}},\ \bibinfo
  {pages} {07D126} (\bibinfo {year} {2012}{\natexlab{b}})}\BibitemShut
  {NoStop}%
\bibitem [{\citenamefont {M{\'e}ndez}\ \emph {et~al.}(2011)\citenamefont
  {M{\'e}ndez}, \citenamefont {Horsthemke}, \citenamefont {Mestres},\ and\
  \citenamefont {Campos}}]{mendez2011instabilities}%
  \BibitemOpen
  \bibfield  {author} {\bibinfo {author} {\bibfnamefont {V.}~\bibnamefont
  {M{\'e}ndez}}, \bibinfo {author} {\bibfnamefont {W.}~\bibnamefont
  {Horsthemke}}, \bibinfo {author} {\bibfnamefont {P.}~\bibnamefont {Mestres}},
  \ and\ \bibinfo {author} {\bibfnamefont {D.}~\bibnamefont {Campos}},\
  }\href@noop {} {\bibfield  {journal} {\bibinfo  {journal} {Physical Review
  E}\ }\textbf {\bibinfo {volume} {84}},\ \bibinfo {pages} {041137} (\bibinfo
  {year} {2011})}\BibitemShut {NoStop}%
\bibitem [{\citenamefont {M{\'e}ndez}\ \emph {et~al.}(2014)\citenamefont
  {M{\'e}ndez}, \citenamefont {Denisov}, \citenamefont {Campos},\ and\
  \citenamefont {Horsthemke}}]{mendez2014role}%
  \BibitemOpen
  \bibfield  {author} {\bibinfo {author} {\bibfnamefont {V.}~\bibnamefont
  {M{\'e}ndez}}, \bibinfo {author} {\bibfnamefont {S.~I.}\ \bibnamefont
  {Denisov}}, \bibinfo {author} {\bibfnamefont {D.}~\bibnamefont {Campos}}, \
  and\ \bibinfo {author} {\bibfnamefont {W.}~\bibnamefont {Horsthemke}},\
  }\href@noop {} {\bibfield  {journal} {\bibinfo  {journal} {Physical Review
  E}\ }\textbf {\bibinfo {volume} {90}},\ \bibinfo {pages} {012116} (\bibinfo
  {year} {2014})}\BibitemShut {NoStop}%
\bibitem [{\citenamefont {Ma}\ and\ \citenamefont
  {Dudarev}(2011)}]{ma2011langevin}%
  \BibitemOpen
  \bibfield  {author} {\bibinfo {author} {\bibfnamefont {P.-W.}\ \bibnamefont
  {Ma}}\ and\ \bibinfo {author} {\bibfnamefont {S.~L.}\ \bibnamefont
  {Dudarev}},\ }\href@noop {} {\bibfield  {journal} {\bibinfo  {journal}
  {Physical Review B}\ }\textbf {\bibinfo {volume} {83}},\ \bibinfo {pages}
  {134418} (\bibinfo {year} {2011})}\BibitemShut {NoStop}%
\bibitem [{\citenamefont {Thibaudeau}\ \emph {et~al.}(2015)\citenamefont
  {Thibaudeau}, \citenamefont {Tranchida},\ and\ \citenamefont
  {Nicolis}}]{thibaudeau2015non}%
  \BibitemOpen
  \bibfield  {author} {\bibinfo {author} {\bibfnamefont {P.}~\bibnamefont
  {Thibaudeau}}, \bibinfo {author} {\bibfnamefont {J.}~\bibnamefont
  {Tranchida}}, \ and\ \bibinfo {author} {\bibfnamefont {S.}~\bibnamefont
  {Nicolis}},\ }\href@noop {} {\bibfield  {journal} {\bibinfo  {journal} {arXiv
  preprint arXiv:1511.01693}\ } (\bibinfo {year} {2015})}\BibitemShut {NoStop}%
\bibitem [{\citenamefont {Pajda}\ \emph {et~al.}(2001)\citenamefont {Pajda},
  \citenamefont {Kudrnovsk{\`y}}, \citenamefont {Turek}, \citenamefont
  {Drchal},\ and\ \citenamefont {Bruno}}]{pajda2001ab}%
  \BibitemOpen
  \bibfield  {author} {\bibinfo {author} {\bibfnamefont {M.}~\bibnamefont
  {Pajda}}, \bibinfo {author} {\bibfnamefont {J.}~\bibnamefont
  {Kudrnovsk{\`y}}}, \bibinfo {author} {\bibfnamefont {I.}~\bibnamefont
  {Turek}}, \bibinfo {author} {\bibfnamefont {V.}~\bibnamefont {Drchal}}, \
  and\ \bibinfo {author} {\bibfnamefont {P.}~\bibnamefont {Bruno}},\
  }\href@noop {} {\bibfield  {journal} {\bibinfo  {journal} {Physical Review
  B}\ }\textbf {\bibinfo {volume} {64}},\ \bibinfo {pages} {174402} (\bibinfo
  {year} {2001})}\BibitemShut {NoStop}%
\bibitem [{\citenamefont {Lounis}\ and\ \citenamefont
  {Dederichs}(2010)}]{lounis2010mapping}%
  \BibitemOpen
  \bibfield  {author} {\bibinfo {author} {\bibfnamefont {S.}~\bibnamefont
  {Lounis}}\ and\ \bibinfo {author} {\bibfnamefont {P.~H.}\ \bibnamefont
  {Dederichs}},\ }\href@noop {} {\bibfield  {journal} {\bibinfo  {journal}
  {Physical Review B}\ }\textbf {\bibinfo {volume} {82}},\ \bibinfo {pages}
  {180404} (\bibinfo {year} {2010})}\BibitemShut {NoStop}%
\bibitem [{\citenamefont {Levermore}(1996)}]{Levermore:1996ys}%
  \BibitemOpen
  \bibfield  {author} {\bibinfo {author} {\bibfnamefont {C.~D.}\ \bibnamefont
  {Levermore}},\ }\href@noop {} {\bibfield  {journal} {\bibinfo  {journal}
  {Journal of Statistical Physics}\ }\textbf {\bibinfo {volume} {83}},\
  \bibinfo {pages} {1021} (\bibinfo {year} {1996})}\BibitemShut {NoStop}%
\bibitem [{\citenamefont {Eu}(1998)}]{Eu:1998rc}%
  \BibitemOpen
  \bibfield  {author} {\bibinfo {author} {\bibfnamefont {B.~C.}\ \bibnamefont
  {Eu}},\ }\href {\doibase 10.1007/978-94-017-2438-8} {\emph {\bibinfo {title}
  {Nonequilibrium Statistical Mechanics : Ensemble Method}}},\ edited by\
  \bibinfo {editor} {\bibfnamefont {A.}~\bibnamefont {van~der Merwe}},\
  \bibinfo {series} {Fundamental Theories of Physics}, Vol.~\bibinfo {volume}
  {93}\ (\bibinfo  {publisher} {Kluver Academic Publishers},\ \bibinfo
  {address} {Dordrecht},\ \bibinfo {year} {1998})\BibitemShut {NoStop}%
\bibitem [{\citenamefont {Bobryk}(2011)}]{bobryk2011closure}%
  \BibitemOpen
  \bibfield  {author} {\bibinfo {author} {\bibfnamefont {R.~V.}\ \bibnamefont
  {Bobryk}},\ }\href@noop {} {\bibfield  {journal} {\bibinfo  {journal}
  {Physical Review E}\ }\textbf {\bibinfo {volume} {83}},\ \bibinfo {pages}
  {057701} (\bibinfo {year} {2011})}\BibitemShut {NoStop}%
\bibitem [{\citenamefont {Nicolis}\ and\ \citenamefont
  {Nicolis}(1998)}]{nicolis1998closing}%
  \BibitemOpen
  \bibfield  {author} {\bibinfo {author} {\bibfnamefont {C.}~\bibnamefont
  {Nicolis}}\ and\ \bibinfo {author} {\bibfnamefont {G.}~\bibnamefont
  {Nicolis}},\ }\href@noop {} {\bibfield  {journal} {\bibinfo  {journal}
  {Physical Review E}\ }\textbf {\bibinfo {volume} {58}},\ \bibinfo {pages}
  {4391} (\bibinfo {year} {1998})}\BibitemShut {NoStop}%
\bibitem [{\citenamefont {Fox}(1986{\natexlab{a}})}]{fox1986functional}%
  \BibitemOpen
  \bibfield  {author} {\bibinfo {author} {\bibfnamefont {R.~F.}\ \bibnamefont
  {Fox}},\ }\href@noop {} {\bibfield  {journal} {\bibinfo  {journal} {Physical
  Review A}\ }\textbf {\bibinfo {volume} {33}},\ \bibinfo {pages} {467}
  (\bibinfo {year} {1986}{\natexlab{a}})}\BibitemShut {NoStop}%
\bibitem [{\citenamefont {Fox}(1986{\natexlab{b}})}]{fox1986uniform}%
  \BibitemOpen
  \bibfield  {author} {\bibinfo {author} {\bibfnamefont {R.~F.}\ \bibnamefont
  {Fox}},\ }\href@noop {} {\bibfield  {journal} {\bibinfo  {journal} {Physical
  Review A}\ }\textbf {\bibinfo {volume} {34}},\ \bibinfo {pages} {4525}
  (\bibinfo {year} {1986}{\natexlab{b}})}\BibitemShut {NoStop}%
\bibitem [{\citenamefont {Ramirez-Piscina}\ and\ \citenamefont
  {Sancho}(1988)}]{ramirez1988higher}%
  \BibitemOpen
  \bibfield  {author} {\bibinfo {author} {\bibfnamefont {L.}~\bibnamefont
  {Ramirez-Piscina}}\ and\ \bibinfo {author} {\bibfnamefont {J.~M.}\
  \bibnamefont {Sancho}},\ }\href@noop {} {\bibfield  {journal} {\bibinfo
  {journal} {Physical Review A}\ }\textbf {\bibinfo {volume} {37}},\ \bibinfo
  {pages} {4469} (\bibinfo {year} {1988})}\BibitemShut {NoStop}%
\bibitem [{\citenamefont {Furutsu}(1963)}]{Furutsu:1963sf}%
  \BibitemOpen
  \bibfield  {author} {\bibinfo {author} {\bibfnamefont {K.}~\bibnamefont
  {Furutsu}},\ }\href@noop {} {\bibfield  {journal} {\bibinfo  {journal}
  {Journal of Research of the National Bureau of Standards}\ }\bibinfo {series}
  {D},\ \textbf {\bibinfo {volume} {67D}},\ \bibinfo {pages} {39} (\bibinfo
  {year} {1963})}\BibitemShut {NoStop}%
\bibitem [{\citenamefont {Novikov}(1965)}]{Novikov:1965fp}%
  \BibitemOpen
  \bibfield  {author} {\bibinfo {author} {\bibfnamefont {E.~A.}\ \bibnamefont
  {Novikov}},\ }\href@noop {} {\bibfield  {journal} {\bibinfo  {journal}
  {Soviet Physics JETP}\ }\textbf {\bibinfo {volume} {20}},\ \bibinfo {pages}
  {1290} (\bibinfo {year} {1965})}\BibitemShut {NoStop}%
\bibitem [{\citenamefont {Venkatesh}\ and\ \citenamefont
  {Patnaik}(1993)}]{Venkatesh:1993kb}%
  \BibitemOpen
  \bibfield  {author} {\bibinfo {author} {\bibfnamefont {T.~G.}\ \bibnamefont
  {Venkatesh}}\ and\ \bibinfo {author} {\bibfnamefont {L.~M.}\ \bibnamefont
  {Patnaik}},\ }\href@noop {} {\bibfield  {journal} {\bibinfo  {journal}
  {Physical Review E}\ }\textbf {\bibinfo {volume} {48}},\ \bibinfo {pages}
  {2402} (\bibinfo {year} {1993})}\BibitemShut {NoStop}%
\bibitem [{\citenamefont {San~Miguel}\ and\ \citenamefont
  {Sancho}(1980)}]{san_miguel_fokker-planck_1980}%
  \BibitemOpen
  \bibfield  {author} {\bibinfo {author} {\bibfnamefont {M.}~\bibnamefont
  {San~Miguel}}\ and\ \bibinfo {author} {\bibfnamefont {J.~M.}\ \bibnamefont
  {Sancho}},\ }\href {\doibase 10.1016/0375-9601(80)90579-4} {\bibfield
  {journal} {\bibinfo  {journal} {Physics Letters A}\ }\textbf {\bibinfo
  {volume} {76}},\ \bibinfo {pages} {97} (\bibinfo {year} {1980})}\BibitemShut
  {NoStop}%
\bibitem [{\citenamefont {Lindenberg}\ and\ \citenamefont
  {West}(1984)}]{lindenberg_finite_1984}%
  \BibitemOpen
  \bibfield  {author} {\bibinfo {author} {\bibfnamefont {K.}~\bibnamefont
  {Lindenberg}}\ and\ \bibinfo {author} {\bibfnamefont {B.~J.}\ \bibnamefont
  {West}},\ }\href {\doibase 10.1016/0378-4371(84)90080-3} {\bibfield
  {journal} {\bibinfo  {journal} {Physica A: Statistical Mechanics and its
  Applications}\ }\textbf {\bibinfo {volume} {128}},\ \bibinfo {pages} {25}
  (\bibinfo {year} {1984})}\BibitemShut {NoStop}%
\bibitem [{\citenamefont {Hanggi}\ \emph {et~al.}(1985)\citenamefont {Hanggi},
  \citenamefont {Mroczkowski}, \citenamefont {Moss},\ and\ \citenamefont
  {McClintock}}]{Hanggi:1985fv}%
  \BibitemOpen
  \bibfield  {author} {\bibinfo {author} {\bibfnamefont {P.}~\bibnamefont
  {Hanggi}}, \bibinfo {author} {\bibfnamefont {T.~J.}\ \bibnamefont
  {Mroczkowski}}, \bibinfo {author} {\bibfnamefont {F.}~\bibnamefont {Moss}}, \
  and\ \bibinfo {author} {\bibfnamefont {P.~V.~E.}\ \bibnamefont
  {McClintock}},\ }\href@noop {} {\bibfield  {journal} {\bibinfo  {journal}
  {Physical Review A}\ }\textbf {\bibinfo {volume} {32}},\ \bibinfo {pages}
  {695} (\bibinfo {year} {1985})}\BibitemShut {NoStop}%
\bibitem [{\citenamefont {Blanes}\ \emph {et~al.}(2009)\citenamefont {Blanes},
  \citenamefont {Casas}, \citenamefont {Oteo},\ and\ \citenamefont
  {Ros}}]{Blanes:2009xw}%
  \BibitemOpen
  \bibfield  {author} {\bibinfo {author} {\bibfnamefont {S.}~\bibnamefont
  {Blanes}}, \bibinfo {author} {\bibfnamefont {F.}~\bibnamefont {Casas}},
  \bibinfo {author} {\bibfnamefont {J.~A.}\ \bibnamefont {Oteo}}, \ and\
  \bibinfo {author} {\bibfnamefont {J.}~\bibnamefont {Ros}},\ }\href@noop {}
  {\bibfield  {journal} {\bibinfo  {journal} {Physics Reports}\ }\textbf
  {\bibinfo {volume} {470}},\ \bibinfo {pages} {151} (\bibinfo {year}
  {2009})}\BibitemShut {NoStop}%
\end{thebibliography}%
\end{document}